\documentclass[iop]{emulateapj} 

\usepackage{natbib}
\usepackage{float}
\usepackage{graphicx}

\received{}
\accepted{}
\shorttitle{Parallax to SS CMa}
\shortauthors{Casertano et al.}

\newcommand{\ms}{\hbox{\, m\ s$^{-1}$}} 
\newcommand{\kms}{\hbox{$ \, \rm km\, s^{-1}$}} 
\newcommand{\kmsmpc}{\hbox{$ \, \rm km\, s^{-1} \, Mpc^{-1}$}}

\newcommand{\bq}{\begin{equation}} 
\newcommand{\eq}{\end{equation}}   
\newcommand{\hubbleconst}{\hbox{$H_0$}}

\newcommand{\beq}{\begin{equation}}
\newcommand{\eeq}{\end{equation}}
\newcommand{\beqa}{\begin{eqnarray}}
\newcommand{\eeqa}{\end{eqnarray}}

\newcommand{\PL}{$P\hbox{--}L$}

\newcommand{\muas}{\hbox{$\, \mu\rm as$}}
\newcommand{\mpix}{\hbox{$\, \rm mpix$}}
\newcommand{\mas}{\hbox{$\, \rm mas$}}
\newcommand{\kpc}{\hbox{$\, \rm kpc$}}
\newcommand{\lcdm}{\hbox{$\Lambda$CDM}}

\newcommand{\Filter}[1]{\hbox{\sl #1}}

\long\def\check#1{}
\long\def\hide#1{}
\def\sscma{SS CMa}
\def\syaur{SY Aur}
\def\HST{{\it HST}}

\begin{document} 

\title{Parallax of Galactic Cepheids from Spatially Scanning \\ 
the Wide Field Camera 3 on the Hubble Space Telescope: The Case of SS
Canis Majoris}

\author{Stefano Casertano\altaffilmark{1,2,3}, Adam
G.~Riess\altaffilmark{2,1}, \\
Jay Anderson\altaffilmark{1}, Richard I.~Anderson\altaffilmark{2,4}, J.~Bradley Bowers\altaffilmark{2}, 
Kelsey I.~Clubb\altaffilmark{5}, Aviv R.~Cukierman\altaffilmark{2,6}, \\
Alexei V.~Filippenko\altaffilmark{5}, Melissa L.~Graham\altaffilmark{5}, 
John W.~MacKenty\altaffilmark{1}, Carl Melis\altaffilmark{7}, \\
Brad E.~Tucker\altaffilmark{5}, and Gautam Upadhya\altaffilmark{1,8}}

\altaffiltext{1}{Space Telescope Science Institute, 3700 San Martin
  Drive, Baltimore, MD 21218, USA}
\altaffiltext{2}{Department of Physics and Astronomy, Johns Hopkins
  University, Baltimore, MD 21218, USA}
\altaffiltext{3}{email address: stefano@stsci.edu.}
\altaffiltext{4}{D\'epartement d'Astronomie, Universit\'e de Gen\`eve, 
  Ch.~des Mailletes 51, CH-1290 Sauverny, Switzerland}
\altaffiltext{5}{Department of Astronomy, University of California,
  Berkeley, CA 94720-3411, USA}
\altaffiltext{6}{Department of Physics, Stanford University, Stanford, CA 94305-4060, USA.}
\altaffiltext{7}{Center for Astrophysics and Space Sciences, University of California,
  San Diego, La Jolla, CA 92093-0424, USA}
\altaffiltext{8}{Department of Physics, University of Chicago, Chicago, IL 60637, USA}
  
\begin{abstract} 

We present a high-precision measurement of the parallax for the 12-day
Cepheid SS Canis Majoris, obtained via spatial scanning with the Wide
Field Camera 3 (WFC3) on the {\it Hubble Space Telescope (HST)}.
Spatial scanning enables astrometric measurements with a precision of
$ 20 \hbox{--} 40 \muas $, an order of magnitude better than pointed
observations.  SS CMa is the second Cepheid targeted for parallax measurement
with {\HST}, and is the first of a sample of eighteen long-period ($ \gtrsim
10 $ days) Cepheids selected in order to improve the calibration
of their period-luminosity relation
and eventually permit a determination of the Hubble constant
{\hubbleconst} to better than 2\%.  The parallax of SS CMa is found to
be $ 348 \pm 38 \muas $, corresponding to a distance of $ 2.9 \pm
0.3 \kpc $.  We also present a refinement of the static geometric distortion
of WFC3 obtained using spatial scanning observations
of calibration fields, with a typical magnitude $ \lesssim 0.01 $ pixels 
on scales of 100 pixels.
 
\end{abstract} 


\section{Introduction} 

A precise test of the cosmological model can be performed by combining
present cosmic microwave background (CMB) measurements \citep{bennett13, planckpaperxiii} with a
percent-level determination of the local Hubble constant
{\hubbleconst} \citep{hu05}.  More than 70 years of work from \citet{hubble29}
through the first decade of observations with the {\it Hubble Space
Telescope (HST)} have resulted in a $\sim 10$\% measurement of
{\hubbleconst} \citep{freedman01,sandage06}, with much of the
remaining uncertainty being of a systematic nature.  \citet{riess11}
sharply reduced the uncertainty to 3.3\%, to a value of $ 73.8 \pm 2.4
\kmsmpc $, thanks to four improvements in the distance ladder
consisting of Cepheids and Type Ia supernovae (SNe~Ia): (1) calibrating
eight modern SNe~Ia with Cepheids, (2) observing Cepheids in the
near-infrared (NIR) to reduce the impact of extinction and
metallicity, (3) the use of two new geometric calibrations of
Cepheids---parallaxes of Galactic Cepheids from the {\it HST\/} Fine
Guidance Sensor \citep[FGS;][]{benedict07} and the 3\% geometric maser
distance to NGC 4258 \citep[][and references therein]{humphreys13},
and (4) calibrating all extragalactic Cepheid photometry with a single
camera, WFC3, to remove cross-instrument zeropoint errors.

While local determinations of {\hubbleconst} place it in the range of
$ 70 \hbox{--} 75 \kmsmpc $ (see, e.g., the reviews by \citealt
{livio13} and \citealt {freedman10}), the predictions from CMB measurements
with a {\lcdm} cosmology find a range
of $ 67 \hbox{--} 70 \kmsmpc $ \citep{bennett14, planckpaperxiii}, indicating tension
between the two sets of determinations.  \citet{addison15} carries out
a comparative reanalysis of {\it Planck} and {\it WMAP} data \citep[see also][and
references therein]{bennett13, planckpaperxiii}, finding that the
{\it Planck} measurements below $ \ell \sim 1000 $ are consistent with {\it WMAP},
while higher multipoles may be inconsistent.  The apparent discrepancy
between local measurements of {{\hubbleconst}} and the values
predicted from cosmological results may indicate deviation from the
{\lcdm} model or new physics \citep[see, e.g.,][]{wyman14}, although
\citet{bennett14} find that the evidence for a discrepancy is
inconclusive.  A resolution on the origin and magnitude of this
potential tension is best found in improving the measurements
themselves, especially those at low redshift, which have a larger
statistical uncertainty.

\begin{figure}[ht]
\includegraphics[width=\columnwidth,bb=70 70 440 605] {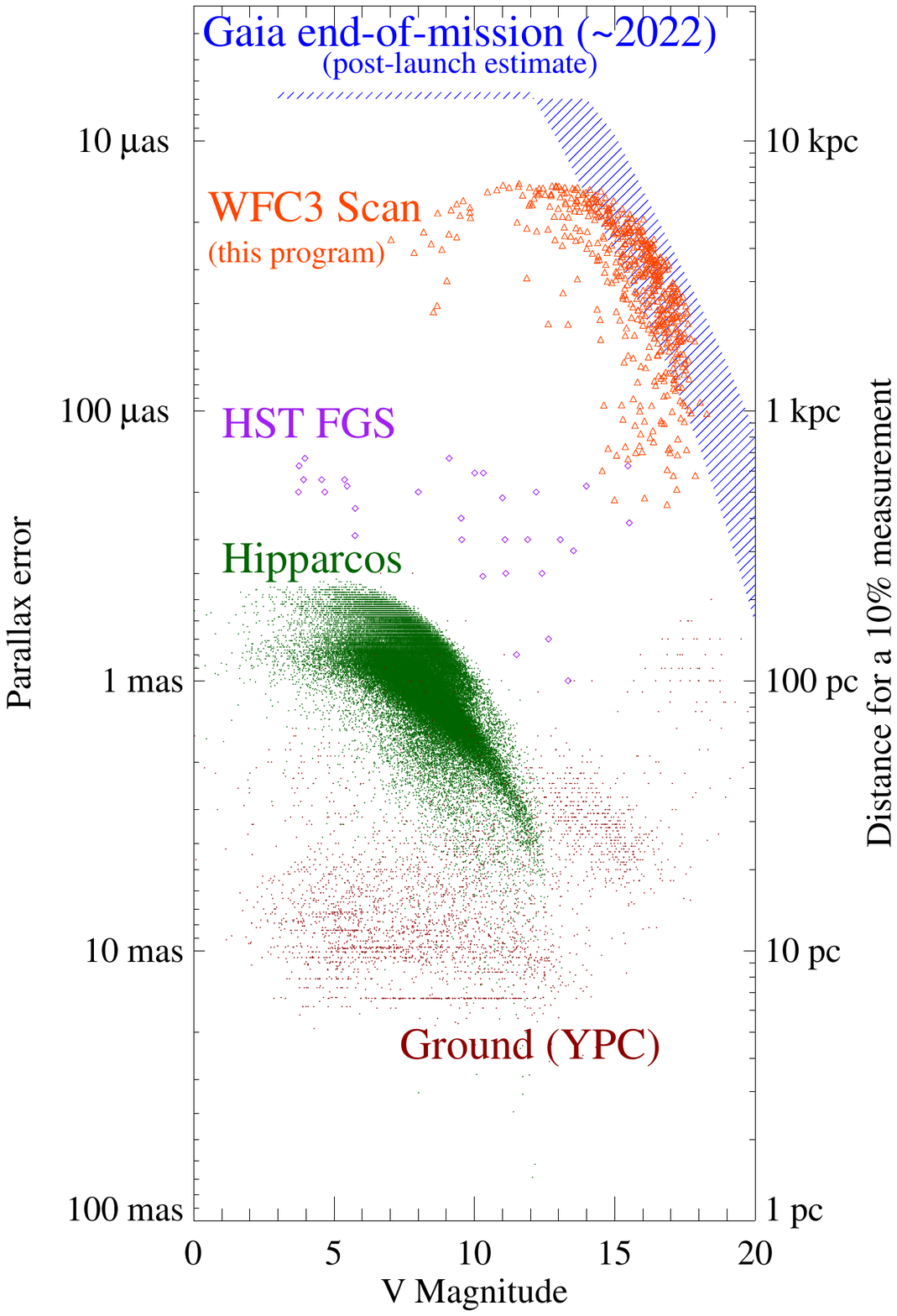}
\caption {\label{fig:parallax_history} Precision of parallax
measurements vs. apparent luminosity from ground and from space,
1995--2022.  The right-hand ordinate axis shows the distance at which the error
exceeds 10\%.  (Brown) Ground-based measurements from the Yale
Parallax Catalog \citep{vanaltena95}.  (Green) stars with a better
than 3-$\sigma$ measurement from {\it Hipparcos} \citep{perryman09}.
(Purple) Measurements based on {\HST}/FGS data \citep{macconnell97, hershey98,
benedict00, benedict01, benedict02, benedict07, benedict09,
benedict11, nelan13}.  (Orange) Projected five-epoch precision for
target and reference stars from the Cepheid fields observed with
{\HST}/WFC3 using spatial scanning.  (Blue) Range of expected precision
for {\it Gaia} observations, according to the post-launch estimates in
\citet{debruijne15}.  With the exception of a few radio-wavelength
measurements \citep{reid14}, only {\HST} spatial scanning and {\it Gaia} can
push the 10\% precision horizon beyond 1 {\kpc}.}
\end{figure}

Starting with \citet{riess09a} and then in \citet{riess11}, we are following a program of
rebuilding the foundation of the local distance ladder by increasing
the range and precision of trigonometric parallax measurements in
order to reach long-period ($ P > 10 $ days) Milky Way Cepheids,
nearly all of which are beyond a distance of $ 2 \kpc $.  In \citet[hereafter Paper 1]{riess14}
we presented a new observational approach to achieve parallax accuracy of
$ \sim 30 \muas $ by spatially scanning the WFC3 camera on {\HST}.  In
principle, this method has the promise of achieving a factor of 10--20 
improvement over conventional pointed observations or FGS measurements
\citep {bellini11, benedict07}.  This method was demonstrated via five
epochs of measurements, spaced every six months, of the field around
SY Aurigae, a 10-day Cepheid for which we reported a parallax of $ 428
\pm 54 \muas $ (statistical).  While confirming the promise of the
method, Paper~1 highlighted several improvements in the experiment
design necessary to achieve the desired measurement precision of $
30\hbox{--}40 \muas $.  Most important of these is the selection of
targets with a greater number of reference stars in the field,
especially those no more than 5~mag fainter than the target
Cepheid.  We included those considerations in our approved follow-up
programs to obtain parallaxes for 18 Galactic Cepheids.  The
observations of these 18 Cepheids, stretching over five to nine
epochs, are now concluding their fifth epoch.  Here we present a
detailed analysis of the results for the first of these targets, the
12-day Cepheid {\sscma} at an expected distance of $\sim 3 \kpc $.
The expected precision of the parallax measurement for the target
Cepheids and the reference stars in their fields are shown in
Figure~\ref{fig:parallax_history}.

In December 2013, the European Space Agency launched the mission {\it Gaia}
\citep{prusti12}, which promises to determine the fundamental
astrometric parameters for $\sim 10^9 $ stars in the Galaxy with
unprecedented precision.  Its targets will include hundreds of
Galactic Cepheids, including the targets of our {\HST} program.
End-of-mission results from {\it Gaia}, expected in 2022, are projected to
achieve a parallax precision close to 10 {\muas} for its bright
targets (see Fig.~\ref{fig:parallax_history}), although special
procedures will be needed for targets brighter than $ V \approx 12 $
mag---including most long-period Cepheids close enough to be effective
distance-scale calibrators.  Early reports from the mission indicate
the existence of significant systematic variations of the basic
angle---the separation between the two fields of view $ 106.5^\circ $
apart which lies at the heart of {\it Gaia}'s ability to measure absolute
parallaxes---on periods close to the satellite spin period \citep
{mora14}.  We are optimistic that internal calibrations will enable a full
correction for these variations and the eventual achievement of the
full expected mission precision shown in
Figure~\ref{fig:parallax_history} (see, e.g., \citealt{michalik15}).
Nonetheless, the availability of an external calibration of
comparable, if slightly coarser, precision may also provide a useful
verification of the {\it Gaia} measurements.  Assuming that {\it Gaia} achieves
its stated goals, the calibration of the {\PL} relation for Galactic
Cepheids will likely be better than 1\% in distance, and provide the
ideal anchor for a measurement of the local value of {\hubbleconst}
with unprecedented precision.

The organization of this paper is as follows.  In \S~2 we describe the
refinements since Paper~1 in the use of spatial scanning data to
measure high precision, relative astrometry at a single epoch.  We
also include a description of the calibration observations we have
obtained to improve knowledge of the geometric distortion and
other instrumental properties of WFC3/UVIS.  Section~3 presents the
spectroscopic and photometric data we obtained to characterize the
properties of the reference stars.  We describe in \S~4 refinements in
the algorithms used to combine multiple epochs of spatial scan data to
measure time-dependent astrometry, and we discuss the parallax
measurement thus obtained.  In \S~5 we show how radial-velocity
information can be used to obtain bounds on the effect that binarity
can have on parallax measurements.  Section~6 briefly discusses the
implications of the present and upcoming measurements.

\section{MW Cepheid Parallaxes: A Sample of 18 Targets.}

In Paper 1 we presented our first parallax measurement for a Galactic
Cepheid with WFC3 spatial scans, the case of {\syaur}.  These
observations probed for the first time the stability and accuracy of
the {\HST} focal plane geometry well below the milli-arcsecond (mas) level.
Until our scanned observations, the practical limit of relative
astrometry with WFC3/UVIS was about 0.01 pixels, or {0.4 \mas}
\citep{bellini11}; test data indicated that scanned
observations of bright stars over 1000--4000 pixels had the potential
to achieve a parallax precision of 20--40\muas, about 10 times better
than existing measurements.  For {\syaur} we achieved a final parallax
precision of {54\muas} (statistical).  However, we were unable to fully
determine the systematic uncertainty on this measurement, owing to the
paucity of reference stars which limited our ability to determine the
sensitivity of the result to different processing choices.  
In many ways, {\syaur} was a test case, and the strict
requirements of our measurement process were not known
at the start of our first 2-year campaign.

On the basis of the analysis of the {\syaur} results, we have selected
a sample of 18 additional Galactic Cepheids for which we could expect
to obtain parallax measurements with uncertainty $ \sigma =
30\hbox{--}40 \muas $ in order to improve the calibration of the
Cepheid Period-Luminosity (\PL) relation for the determination of the
Hubble constant.  Cepheids in this sample are listed in Table~1, with
some basic properties; the magnitudes in the Table are as reported by
\citet{vanleeuwen07}.  Note that our program includes obtaining
photometry of the target Cepheids with {\HST} in the same filters used
for those in SN~Ia host galaxies, in order to remove any uncertainties
related to differences between ground-based and {\HST} photometric
systems.  The primary considerations in their selection are: (1)
period longer than $ \sim 10 $ days; (2) $ \sim 10 $ reference stars
within the field (scan length $ > 500 $ pixels) within 5--6~mag of
the Cepheid itself; and (3) an estimated distance less than $ 4 \kpc
$ at the 3-$\sigma$ level.

\begin{figure}[ht]
\includegraphics[width=\columnwidth,bb=62 62 730 560] {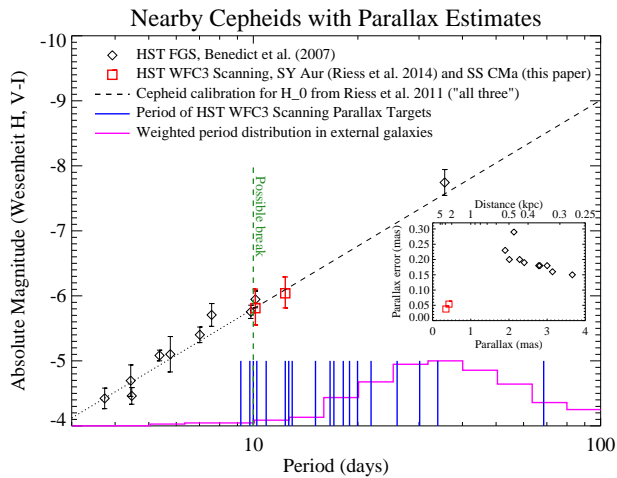}
\caption {\label{fig:plrel} Period-luminosity relation for Galactic
Cepheids with trigonometric distance measurement, and period
distribution for Cepheids used in the $ H_0 $ distance calibration.
The open diamonds are the Galactic Cepheids with \HST-FGS parallax
measurements from \citet{benedict07}, with the Wesenheit absolute
magnitude in the $ H $ band estimated from their distance and uncertainty.
The red squares are {\syaur} and {\sscma}, with the distances determined in Paper~1 and
in this paper, respectively.
The dotted/dashed line shows the {\PL} calibration
obtained in \citet{riess11} when using all three anchors (Galactic
Cepheids, LMC, and NGC 4258).  Because of a possible break at $ P \approx
10 $ days \citep[marked by a vertical green line;][and references therein]{sandage04, ngeow09, kodric15}, 
the line is shown dotted below 10 days and dashed above.  
The magenta histogram indicates the distribution of periods
for Cepheids in SN~Ia hosts (Hoffmann et al.~2016 and Riess et al.~2016, in prep.), scaled to
the same total weight for each host.  The vertical blue bars show the
period of the other Cepheids in our sample (Table~1).  Finally, the
inset shows the parallax and error for the \citet{benedict07} targets
(black diamonds) vs. {\syaur} and {\sscma} (red squares); note that the {\it
absolute} parallax error for {\syaur} and {\sscma} are much smaller than for the
previous targets, but the {\it fractional} parallax error---and thus
the quality of the luminosity calibration---is comparable.}
\end{figure}

The requirement for a period longer than 10 days stems from the desire
to minimize the impact of systematic uncertainties when using the {\PL}
relation to measure the Hubble constant.  Cepheids in external
galaxies, especially the hosts of Type Ia supernovae, can be observed
with adequate accuracy only if they are sufficiently bright, which
implies longer periods.  In practice, most of the information comes
from Cepheids with periods longer than 10 days.  Calibrating the {\PL}
relation with Cepheids of significantly shorter period introduces a
systematic uncertainty related to the slope of the relationship.  In
addition, there are indications that the {\PL} relation has a break in
the neighborhood of 10 days, represented by a change in its slope
\citep[see, e.g.,][and references therein]{sandage04, ngeow09, kodric15}.  If true, this
enhances the reason to use as local calibrators primarily
longer-period Cepheids, which follow the same {\PL} relation as the
Cepheids in supernova host galaxies.  Figure~\ref{fig:plrel} shows the
period-luminosity relation for Galactic Cepheids with measured
parallaxes, the distribution of periods for the Cepheids in supernova
host galaxies (Riess et al.~2016 and Hoffmann et al.~2016, in prep.), and the periods of the
Cepheids in our sample.

As shown by our experience with {\syaur}, bright reference stars are
critical to constrain the relative orientation and variable geometric
transformation between scanning mode exposures.  Shallow
exposures---typically either in narrow-band filters or with the
telescope moving faster than $ 1 \arcsec \, \rm s^{-1} $---are needed
to observe the $ V \approx 9 $~mag Cepheid without saturation.  Deep
exposures in a broad-band filter are needed in order to measure enough
reference stars ($ V < 17 $~mag) to provide a well-constrained absolute
parallax, as discussed in \S~4.  Shallow and deep scanning
observations are obtained within the same orbit, but the analysis of
{\syaur} data shows that the geometric distortion varies enough within
a single orbit, due to the {\HST} day-night cycle, that a second-order polynomial term is needed to
account for its change.  Unless about 10 or more stars are available to
determine the second-order polynomial correction, the uncertainty from
the correction dominates the uncertainty in the Cepheid measurement.
Therefore we require that at least 10 stars be observable 
in the shallow exposures with 
signal-to-noise ratio $ > 30 $ per pixel, which implies stars no more
than 5~mag fainter than the Cepheid, and scan length of at least 500 pixels.

Finally, the requirement on estimated distance ensures that a nominal
error of $ \sim 30 \muas $ in parallax translate into a $ \sim 10\% $ distance
error for each target.  Assuming that each Cepheid is at a distance
consistent with the current {\PL} calibration, a final parallax error of
30{\muas} for each Cepheid, combined with adequate photometry, would
result in a collective calibration of the {\PL} relation to
approximately 0.04 mag, or $ 2\% $ in distance, a significant improvement over
the 3\% uncertainty of the NGC~4258 calibration \citep{humphreys13}.

\subsection {HST Observations}

For each Cepheid, we obtain {\HST} observations in five to nine epochs
at 6 month intervals, ensuring that the
observations are always executed at orientations $ 180^\circ$
apart to within the {\HST} pointing precision (about $ 0.01^\circ$).
The reason is that our measurements are inherently
one-dimensional; we obtain very accurate positions perpendicular to
the scanning direction, and much less accurate (often less so than
direct observations) in the direction along the scan.  In order to
optimally measure the {\it variation} in position of the Cepheid, we
need to ensure that the direction of resolution is always nearly the
same.

To the extent possible, we also need the scan direction to be fixed
with respect to the detector frame, and close to the detector $ Y $
direction.  This minimizes the impact of low-level geometric
distortion, for which only the $ X $ component is needed, and of the
Charge Transfer Efficiency (CTE) effects that are well-documented with
space-based charge-coupled devices \citep[CCDs; see,
e.g.,][]{anderson10}.  In particular, the CTE losses are much smaller
in the detector $ X $ direction \citep{anderson14}; consequently, it
is desirable for the measurement direction to nearly coincide with the
detector $ X $ direction.  (A small angular offset is introduced in
order to vary the pixel phase along the scan.)  The motion of the
Cepheid in the resolution direction is the result of the combination
of the appropriate component of the proper motion and of the parallax
of the target, compared to that of the reference stars.  Ideally, the
date and orientation of each observation should be chosen to maximize
the projection of the parallactic motion along the resolution
direction.  Because of the $ 180^\circ $ change requirement, the date
and allowed orientation range of each observation are constrained,
typically resulting in a projection factor of $ 0.8\hbox{--}0.9 $.  We
allow for a slack of up to one week in the scheduling of each
observation.

At each epoch, we obtain four or five scanned observations.  The first
four are straight scans in the sequence: Forward, deep; Backward,
shallow; Forward, shallow; Backward, deep.  This sequencing helps
average out time variations between deep and shallow scans, which
could otherwise lead to larger systematic differences between deep
scans (for most reference stars) and shallow scans (for the Cepheid
and the brighter reference stars).

If possible, a fifth scan is obtained in so-called ``serpentine''
mode, in which the scan speed is increased to a value sufficient to
avoid saturation of the target Cepheid in the broad-band filter,
typically of order of $ 1\arcsec\hbox{--}4\arcsec \, {\rm s^{-1}} $
(up to an order of magnitude faster than the straight scan).  With
such a high scan speed, the length of the scan in the standard 350 s
exposure time exceeds the size of the detector.  Thus, in order to fit the
length of the scan, it is necessary to ``fold'' the scan itself:
the telescope describes a series of parallel forward and
backward scans, offset by a user-selectable amount in the $ X $
direction.  We use a separation of $ 4 \arcsec $, about 100 pixels.
These scans are more complex to analyze and are more affected by
potential cross-contamination (overlap between scans pertaining to
different stars) because of the higher density of scans, but they
offer the potential for direct comparison of the Cepheid and many of
the reference stars within the same filter, and they can improve the
measurement precision since more pixels are covered.  Discussion of
the analysis of serpentine scans can be found in \S~2.6.1.

As described in Paper~1, in addition to the scanned observations we
also obtain short, pointed observations of the field in order to
determine multi-band photometry of the reference stars in several
medium-band filters, including WFC3's analogs of the Str\"{o}mgren
filters.  The photometry thus obtained, combined with infrared $JHK$
photometry from the Two Micron All-Sky Survey \citep[2MASS;][]{skrutskie06}, 
space-based photometry at 3.6 and $ 4.5 \,\micron $ from 
the {\it Wide-field Infrared Survey Explorer} \citep[{\it WISE};][]{wright10}, 
and ground-based medium-resolution
spectra, is used to obtain spectrophotometric distance estimates of as
many of the reference stars as possible, which is a critical step in
converting the relative parallax measurement for the Cepheid target
into an absolute measurement.  Details on the spectrophotometric data
and distance estimates are in \S~3; the estimate of absolute
parallax is discussed in \S~4.

\subsection {The Case of {\sscma}}

Here we present the results of the analysis for the first of these 18
targets, the fundamental-mode Cepheid SS Canis Majoris (\sscma), with
a period of 12.35 days and a mean magnitude $ \langle V \rangle = 9.9
$ mag.  {\sscma} was identified as variable by
\citet{hoffmeister29}, and a period was determined by
\citet{oosterhof35}.  We have chosen to complete the analysis of
{\sscma} because it is one of the first few Cepheids for which five
epochs of observations have been completed, and because we believe
that it is representative of the possible accuracy of the parallax
measurements for the rest of the targets of our program.

\begin{figure}[ht]
\includegraphics[width=\columnwidth,bb=0 0 455 790] {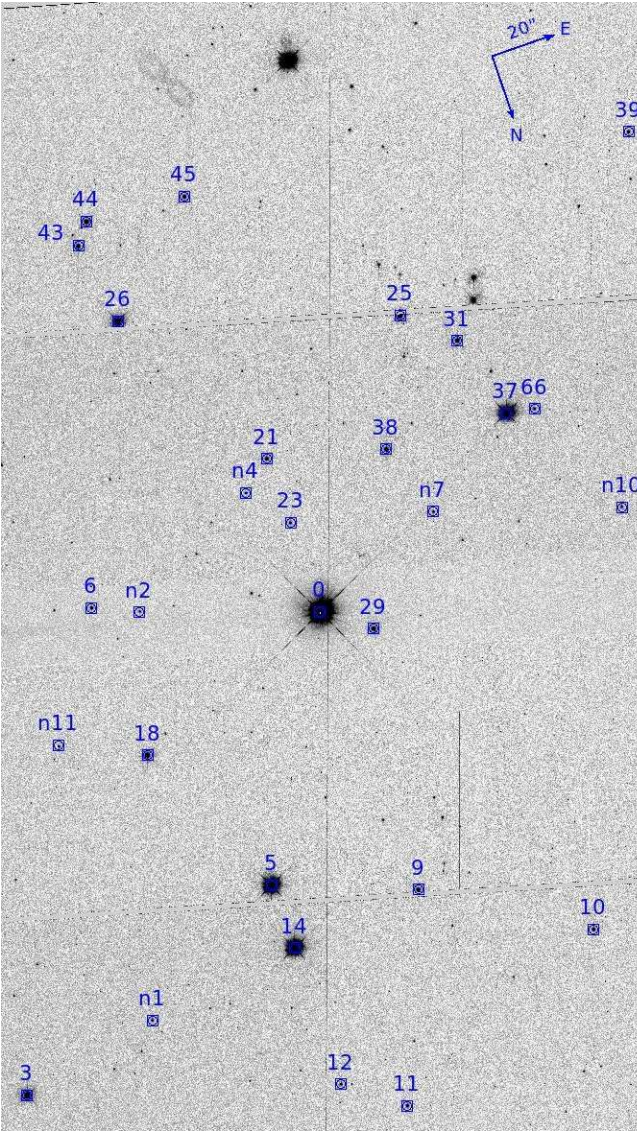}
\caption {\label{fig:direct_image}
Mosaic of direct {\HST} image of the field around {\sscma} in
filter \Filter{F567M}, obtained with WFC3/UVIS in $ 2 \times 2 $ binned mode.  The Cepheid
at the center of the field is saturated.  This image results from a
mosaic of two exposures vertically displaced by $ \sim 20'' $ in order
to cover the full field appearing in the scanned images.  The two
diagonal gaps result from the separation between the WFC3/UVIS
detectors.  Reference stars used in the analysis are marked with the
identification reported in Table~2.  Star~29 is the putative companion of
{\sscma}.}
\end{figure}

\begin{figure}[ht]
\includegraphics[width=\columnwidth,bb=115 56 1142 1078] {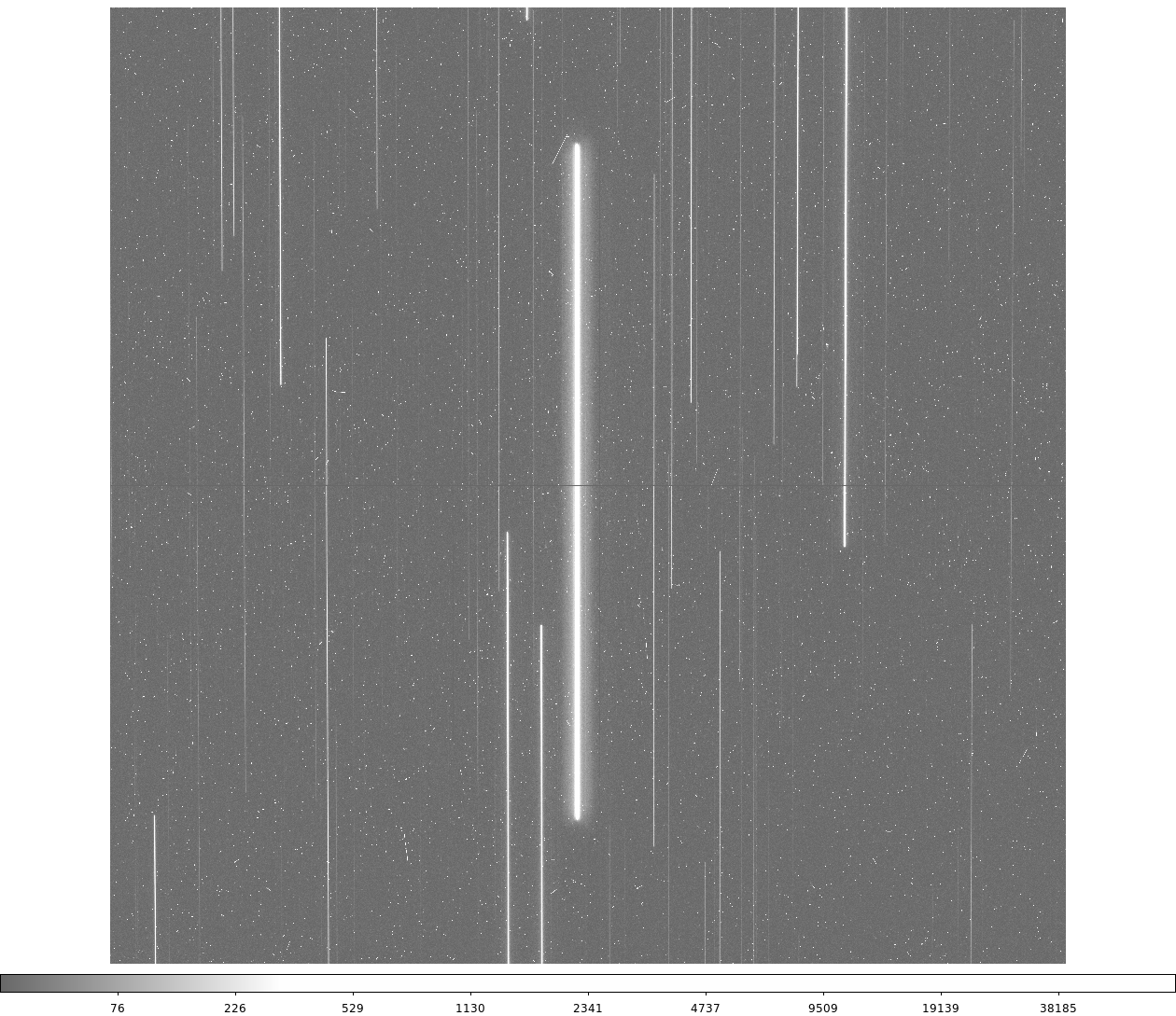}
\vspace {10pt}
\includegraphics[width=\columnwidth,bb=115 56 1142 1078] {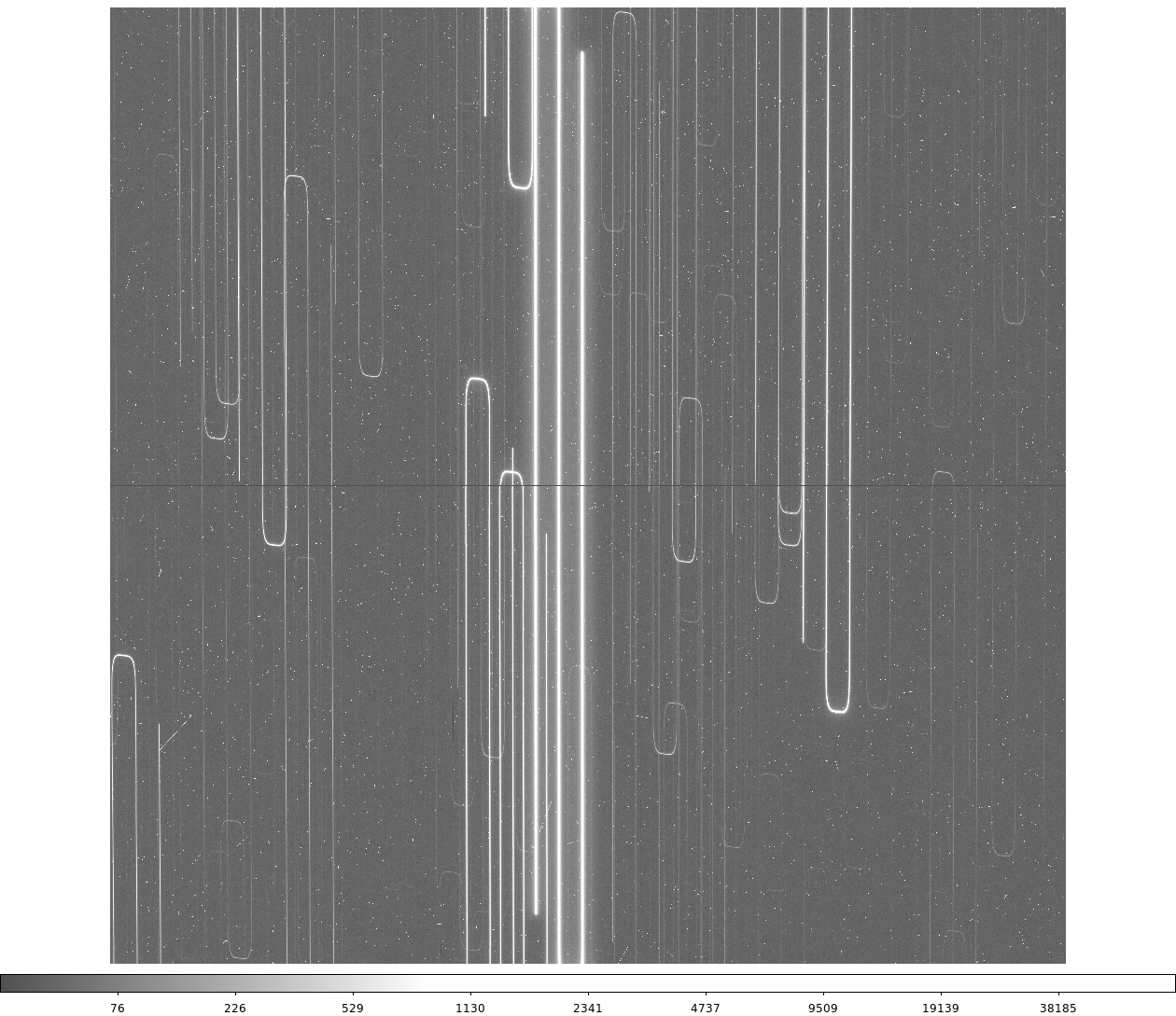}
\caption {\label{fig:scan_image}
(Top) Direct scan of the {\sscma} field in \Filter{F621M}.  The central
strip is the Cepheid, which is not saturated in this image.  (Bottom)
The serpentine scan obtained in \Filter{F606W}.  For this image, the scan speed
of the telescope was sufficiently high that the Cepheid does not saturate,
resulting in a total scan length of over 8000 pixels.  Consequently,
the scan is folded twice, resulting in three separate legs for each
star.}
\end{figure}

Figure~\ref{fig:direct_image} shows a mosaic of the region of sky around
{\sscma} in the filter F547M, as obtained from the very short
observations included in our program.  The area represents
approximately 2 WFC3/UVIS fields of view, stacked vertically with an
overlap of $ \sim 20\arcsec $.  Reference stars and their designations
are indicated.  The Cepheid (Star 0) is saturated.  The diagonal
bands are caused by the gap between the two detectors in WFC3/UVIS; more
observations in the future will help cover this gap.
Figure~\ref{fig:scan_image} shows a normal scan (top) and a serpentine
scan (bottom); serpentine scans are discussed further in \S\S~2.4 and
2.6.1.

The Cepheid {\sscma} has been discussed in the literature as a
potential binary.  \citet{evans94} identify a nearby faint blue star
(Star~29 in Fig.~\ref{fig:direct_image}; 13{\arcsec} from the Cepheid,
$ V \approx 15.51 \hbox{--} 15.58 $ mag) which they argue is likely to be
a physical companion, on the basis of its estimated distance modulus
and of probabilistic arguments.  We will show in \S~4 that
astrometric and spectrophotometric evidence suggests that Star~29 is
significantly closer than the Cepheid, and therefore not a physical
companion.  \citet{szabados96} reports an apparent difference of $
\sim 15 \, \kms $ between the radial velocities (RVs) measured by
\citet{joy37} and \citet{coulson85}.  If interpreted as caused by
binarity, this measurement would indicate a massive companion in an
orbit with period from a few up to 20 years, depending on its
eccentricity.

Binarity can potentially bias the astrometric parallax determination
if the orbital motion of the Cepheid itself, sampled at the times of
the astrometric observations, has a sufficient component to contribute
to the parallax signature.  This is most likely for orbits with period
$ \sim 1 $ year; very long-period binaries will produce primarily a
proper-motion bias, and short-period binaries have a small astrometric
signature that will typically average out in the measurements.

For the case of {\sscma}, we have obtained new RV
measurements, discussed in detail in \S~5, which demonstrate that the
contribution of binary motion compatible with the observations is most
likely below a few {\muas}, thus significantly smaller than
the uncertainty in the parallax measurement.

\subsection{HST Spatial Scans for Astrometry}

In Paper~1, we demonstrated that for sufficiently bright sources,
scanning-mode {\HST} observations, in which the telescope is slewed
during the exposure and each star leaves a trail of light nearly along
a pixel axis, can achieve positional measurements with a precision up
to 0.5-1 millipixels (1 millipixel, or {\mpix}, is about $ 40 \muas
$) in the direction perpendicular to the scan, about a factor of 10
better than optimal measurements from pointed images.  However, we
also discovered that at the millipixel level, several systematic
effects come into play and need to be properly calibrated in order to
fully realize a corresponding {\it accuracy} of $ \lesssim 40 \muas $
in the parallax measurement.

\begin{figure}[h]
\includegraphics[width=\columnwidth,bb=62 62 700 558]{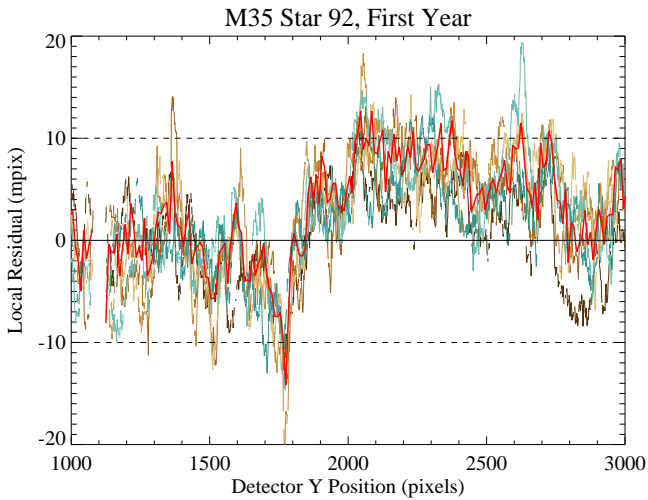}
\caption {\label{fig:delta_geo_stability}
Repeatability of the differential geometric distortion over
a 2-orbit time period.  The thin colored lines show the offset in the
detector $ X $ direction for a single bright star in M35 over repeated
scans, smoothed over 20 pixels, and expressed in millipixels.  The
horizontal dashed lines indicate the nominal precision of the standard
geometric distortion, 0.01 pixels.  The thicker red line is the mean
differential distortion and has an root-mean square (RMS) amplitude of 6.0 mpix;
the RMS difference between individual lines and the mean is 4.0 mpix.
The typical statistical measurement uncertainty for this star 
is $ \approx 2 \mpix $ per smoothed pixel per scan, and under
$ 1 \mpix $ per cell.}
\end{figure}

\begin{figure}[h]
\includegraphics[width=\columnwidth,bb=62 62 700 558] {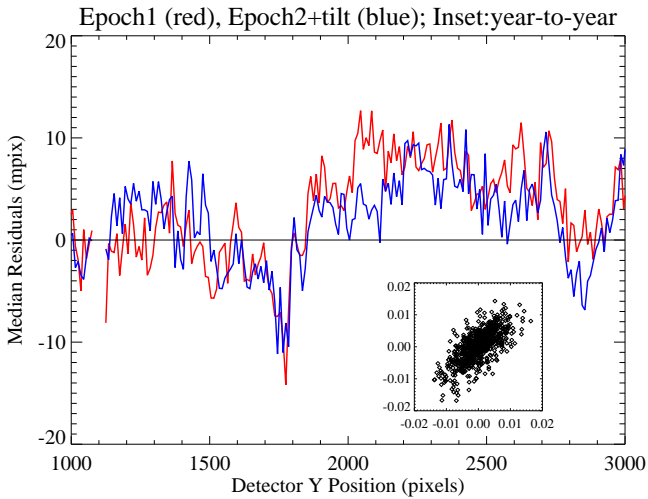}
\caption {\label{fig:delta_geo_secular}
Comparison between the differential geometric distortion
measured for the same star in the same detector location one year
apart.  The red line is the same as in
Figure~\ref{fig:delta_geo_stability}; the blue line is the result of the same
measurement a year later, with a global tilt removed.  The
measurements track one another very closely, demonstrating that the local
differential geometric distortion remains constant over a one-year
period.  The RMS difference between the lines is 3.9 mpix,
corresponding to a repeatability of about 2.8 mpix.  The inset shows the
differential distortion measured in each $ 100 \times 100 $ pixel
cell in Year 1 (abscissa) vs.~Year 2 (ordinate), expressed in pixels.
The year-over-year correlation coefficient is 0.70.}
\end{figure}

Perhaps the most problematic systematic effect is in the
insufficiently characterized, and variable, geometric distortion
solution for the WFC3/UVIS camera.  The mapping between pixels and the
sky is well established and calibrated for direct images, with a
residual uncertainty currently estimated below 0.6\% of a pixel
(root-mean square) over the field \citep{bellini11}---fully adequate
for the astrometric interpretation of direct, pointed images.
However, this uncertainty is an order of magnitude higher than the
requirements of our program.  Calibration observations of the field of
M35, obtained one year apart in 2012 and 2013, show that much of the
residual geometric distortion at the sub-0.01 pixel level is static
and smooth, in that the residuals vary slowly over the field of view
(on scales of $ \sim 100 $ pixels) and are highly correlated from year
to year.  

Figure~\ref{fig:delta_geo_stability} shows that the pattern of
residual geometric distortion is repeatable over short (orbital) time
scales.  Each line shows the variation in the measured $ X $ position
along the scan for the same bright star, located at the same detector
position, in several consecutive scans over a two-orbit period, after
subtracting the jitter pattern for that observation.  To reduce
pixel-to-pixel noise, the lines have been smoothed with a 20-pixel
length.  Without residuals in the geometric distortion or other
disturbance factors, these lines should all be consistent with zero
(i.e., a constant $ X $ position along the scan).  Instead, there is a
definite pattern of deviations, and this pattern is consistent from
scan to scan.  The horizontal dashed lines indicate the nominal
precision of 10 {\mpix} for pointed observations, which is also the
expected accuracy of the standard geometric distortion solution.  This
distortion pattern repeats closely even a year later:
Figure~\ref{fig:delta_geo_secular} shows the median pattern for the
same star in the same sets of observations taken one year apart.
(Consistent with our treatment of all scans, an overall tilt of the
lines has been solved for and subtracted.)  Again, the patterns are
very similar, strongly suggesting that the residual geometric
distortion is stable over time.  The inset in
Figure~\ref{fig:delta_geo_secular} shows the measured differential
geometric distortion in each cell in Year 1 (abscissa) vs.~Year 2
(ordinate); the two quantities are highly correlated ($ r \approx 0.70
$), showing that a significant part of the distortion remains the same 
from year to year over the whole field of view.

However, the differences between scans in the same year, or between
years, is larger than nominal statistical errors, which are 
below $ 1 \mpix $ per cell.  Time-dependent geometric distortion, identified
and discussed in Paper~1 for both calibration and {\syaur}
observations, contribute to these differences.  We thus attempt to
characterize and correct for both a {\it static} and a {\it
time-dependent} component of the geometric distortion correction, in
different ways.

\begin{figure}[ht]
\includegraphics[width=\columnwidth,bb=38 122 573 674] {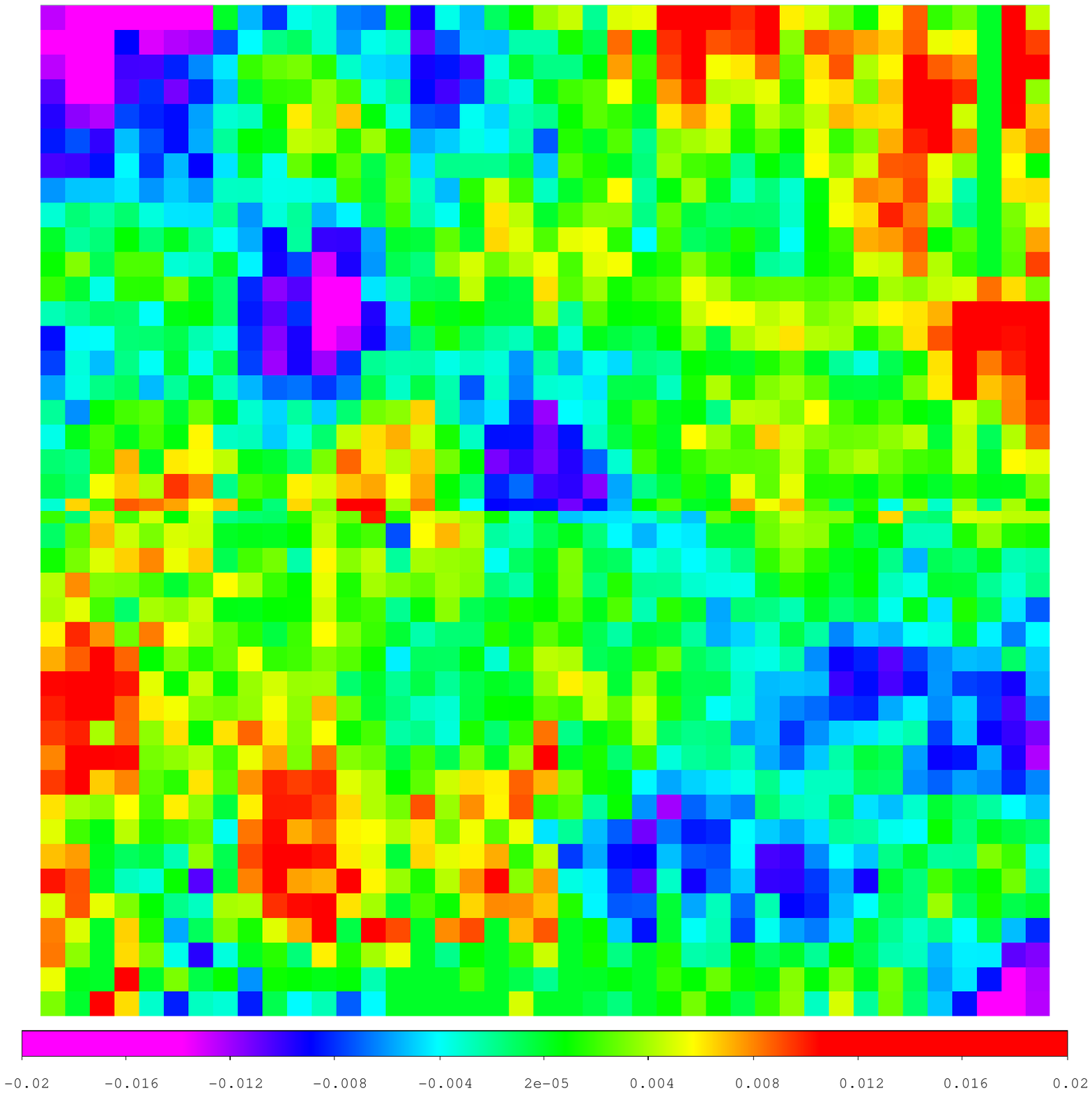}
\vspace{5pt}
\includegraphics[width=\columnwidth,bb=38 122 573 674] {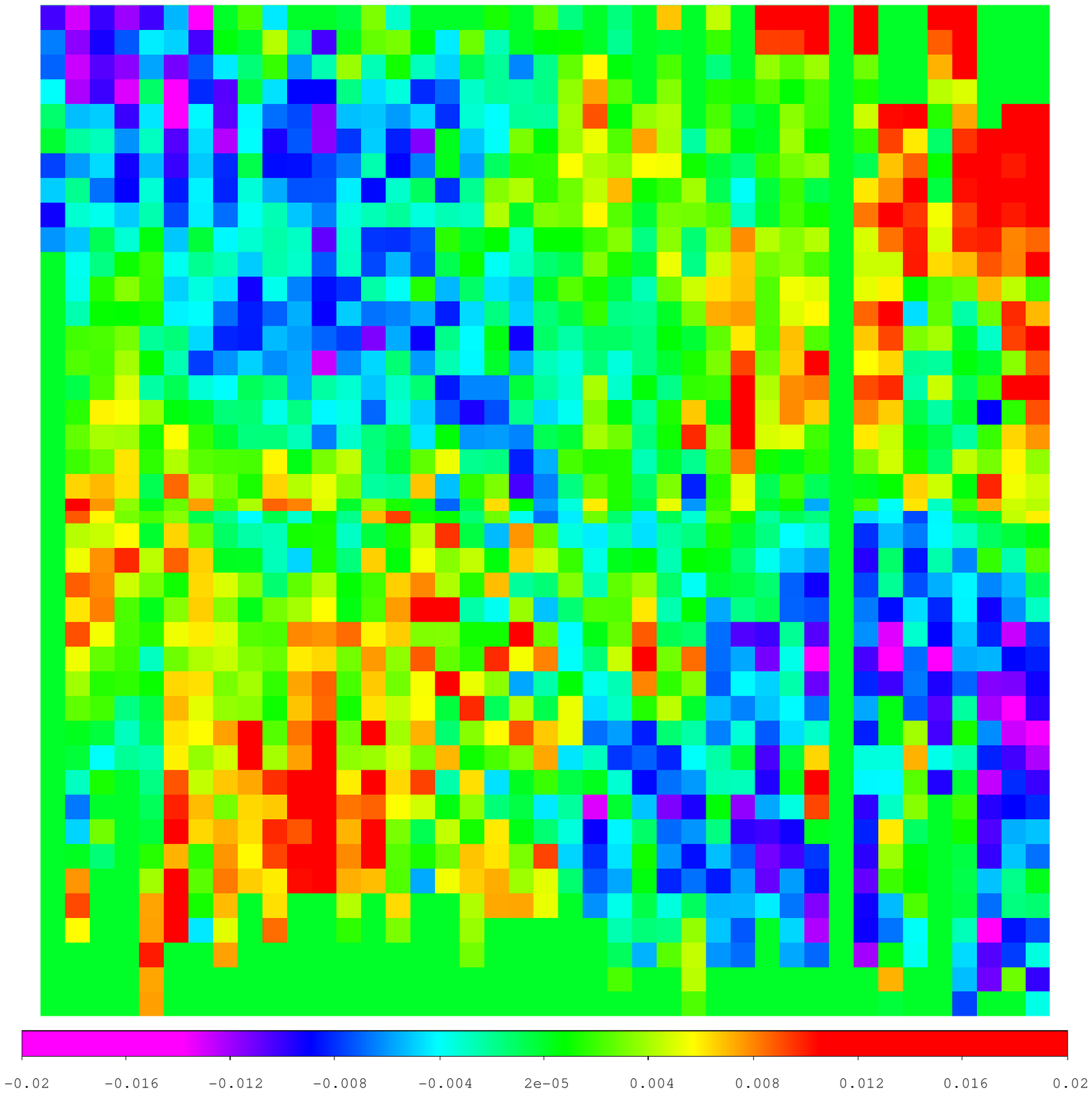}
\vspace{-20pt}
\caption{\label{fig:delta_geo}
(Top) Pseudo-color representation of the static correction to the
default geometric solution in the $ X $ direction for \Filter{F606W}, averaged
over $ 100 \times 100 $ pixel cells.  The correction was obtained from
scanned observations of a field in the open cluster M67.  A total of
20 $ X $ dithers were obtained in order to cover the detector as well
as possible.  The color bar ranges from $ -0.02 $ to 0.02 pixels.  Over
most of the field of view, the correction is less than 0.01 pixels
(blue to orange), but there are small regions with large corrections
(negative: purple, positive: red).  (Bottom) Same for \Filter{F621M}, from 15
dithers in the open cluster M48 to better match the sensitivity of the
narrower filter.  The overall pattern is similar to {\Filter{F606W}} (see also
Fig.~\ref{fig:delta_geo_correlation}).  About 5\% of the cells have no
measurement.}
\end{figure}

\subsubsection {Static Correction to the Geometric Distortion Solution}

Even neglecting the time dependence of the geometric distortion
solution, any residual static term will affect our solution.  The
reason is that parallax observations need to take place approximately
at 6-month intervals, and orbital geometry mandates that observations 6
months apart cannot be taken at the same orientation, although they
can generally be taken at orientations $ 180^\circ $ apart.  Thus, for a
typical target, there will be three epochs taken at one orientation,
and two taken at an orientation different by $ 180^\circ $.  Changing the
telescope roll by $ 180^\circ $ ensures that the resolution direction,
perpendicular to the scan direction and thus typically along the
detector $ X $ direction, remains the same on the sky, thus greatly
simplifying the analysis and improving the accuracy of the results.

Target and reference stars can be placed essentially at the same
detector location for each orientation; as we are interested in {\it
relative variations} in the stars' position, small errors in the
geometric distortion solution, which typically behave smoothly over
the detector, will cancel out.  However, this is not the case across
orientations, when each target's location moves to a completely
different place in the detector, and thus the accuracy in the
geometric distortion solution comes in fully.

\begin{figure}[ht]
\includegraphics[width=\columnwidth,bb=70 70 580 580] {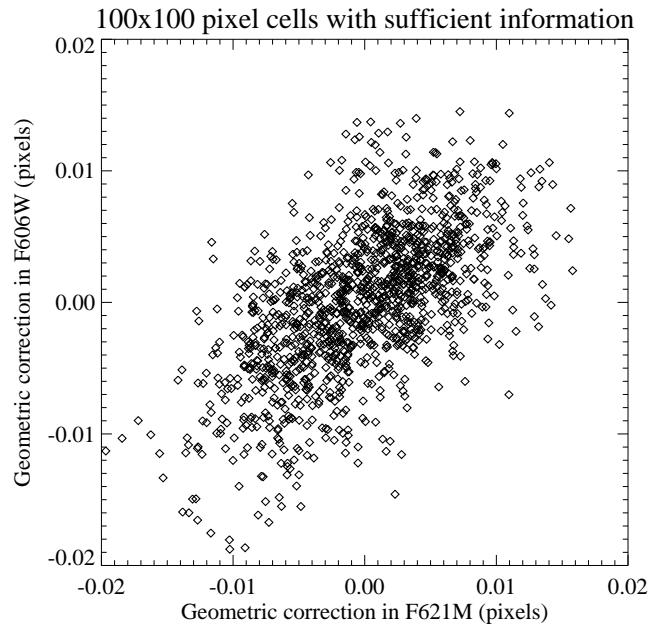}
\caption {\label{fig:delta_geo_correlation}
Comparison between the geometric distortion residuals in
{\Filter{F621M}} and \Filter{F606W}, respectively, for each $ 100 \times 100 $ pixel cell in the
detector.  There is a clear correlation between the residuals,
indicating that the bulk of the correction is in common between
filters.  However, about 30\% of the residual correction, as
determined from these measurements, differs between the filters,
suggesting that a better correction is achieved with a filter-specific
correction.}
\end{figure}

In order to improve the static geometric distortion solution, we have
analyzed calibration observations taken of two open clusters, M48 and
M67, which offer a wealth of bright stars and thus allow a dense
sampling of the detector in as few as 10 dithers.  Such observations
have been obtained as part of the Cycle 22 WFC3 calibration program,
and have demonstrated that a static term, sampled on a grid of $
100\times 100 $ pixel cells, can account for about half of the
deviation from an accurate solution.  (The differences remain 
larger than the statistical measurement errors, which are below $ 1
\mpix $, in part because of the time-dependent correction
discussed in Section~2.3.2.)
We therefore employ this solution as a correction to the
default geometric distortion solution obtained by \citet{bellini11}.
The pattern of static geometric distortion thus obtained is shown in
Figure~\ref{fig:delta_geo}; the top panel is for \Filter{F606W}, and
the bottom panel for \Filter{F621M}.  The two patterns look remarkably
similar; Figure~\ref{fig:delta_geo_correlation} shows the strong
correlation ($ r \approx 0.7 $) between the geometric distortion
corrections thus obtained, despite the fact that the observations
targeted different star fields, in different filters, and were taken
more than a month apart.

\subsubsection {Time-Dependent Correction}

Even for observations obtained with stars at exactly the same location
and in back-to-back exposures, we find that there is a smooth
variation in the relative positions measured for each star, which
generally can be well approximated as a low-order polynomial function
of the position on the detector.  This variation appears correlated
with the nominal focus position of the telescope, which is provided
after the fact by the Space Telescope Science Institute (STScI) Telescopes
group as a result of a temperature-based model of the telescope's 
optical train ({\footnotesize \tt \verb=http://www.stsci.edu/hst/observatory/focus/FocusModel=}).  There
is also a suggestion that the polynomial coefficients are
correlated, and in fact a principal-component analysis (PCA) of
the correction polynomial shows that two or three parameters suffice
to account for over 95\% of the correction.

Although the correlations of the polynomial correction with focus and
internally across coefficients are highly indicative, we do not yet
have sufficient information to characterize the required correction
directly from estimated focus.  We therefore resort to a self-calibration approach, in
which the polynomial distortion is part of the model and is chosen so
as to minimize the source-by-source residuals of the full set of
observations across epochs.

\subsection{Designing the Observations}

Based on our experience observing the field of {\syaur} (Paper~1),
we developed simulation tools to optimize the observations
of other Cepheids and we applied these to the field of {\sscma}.  As in
{\syaur}, we selected positions, scan speeds, scan lengths, and filters
to allow the highest quality parallax measurements from the field
(see Paper~1 for these details).  The field of {\sscma} provided more
than twice as many reference stars of intermediate brightness,
used to register the deep and shallow scans, as the field of {\syaur}.

In addition to the straight scans we discussed in Paper~1---two each
in broad and narrow-band filters at each epoch---we have also obtained
and processed {\it serpentine} scans, in which the telescope moves
through the field in a boustrophedonic pattern (down to up, shift
right, up to down, shift right, etc.), as shown in the bottom panel of
Figure~\ref{fig:scan_image}.  With this pattern scanned at a rate of $
1.5 \arcsec\, {\rm s}^{-1}$, the target Cepheid does not saturate even
in the broad-band filter.  However, at this rate the telescope will
traverse about $ 525 \arcsec $ over a 350~s exposure, almost four times
the WFC3 field of view.  Taking exposures shorter than 350~s is not
desirable because of how memory is managed in WFC3; therefore, in order to keep
the target and most of the reference stars on chip for the largest
fraction of the time, and thus to collect as many photons as possible,
the scan pattern is folded into multiple near-straight scan lines, all
approximately along the $ Y $ direction, with small ``crossbars'' between
them.  In the case of {\sscma}, the {\Filter{F606W}} serpentine scans had three
legs, and the overall length was such that no crossbars are visible
for the Cepheid; the turnaround occurred while the Cepheid was
off-chip.  Owing to planning priorities, no serpentine scan was obtained
in the fifth and final epoch.

The advantage of the serpentine scanning pattern is that the Cepheid
and most of the reference stars are observable in the same exposure;
thus, it is possible to solve directly for relative and absolute
parallaxes without the potential for inconsistencies in the geometric
distortion solution across filters.  The disadvantages are that (1) the
telescope scans faster than in the narrow-band frame with equivalent
count rate, therefore providing less local contrast against the sky
background for faint reference stars (this is partially compensated by
the larger number of points per star); (2) the exposures are more
crowded, with a greater chance that otherwise fine reference stars
will be marred by an overlapping trace from a different star, and (3)
the motion of the telescope is more complex, and it is more difficult
to identify fixed points such as the start and end of each scan.  For
example, for {\sscma} the scan length in the middle leg exceeds the
size of the WFC3 instantaneous field of view, so that neither its
start nor its end are visible.  Thus, it is more challenging to
determine the position of the reference points along each leg, other
than the overall start and end.  These difficulties notwithstanding,
we have been able to process serpentine scans with methods similar to
our other scans, and the results are incorporated in our astrometric
solution (see \S~2.6.1).

\subsection{Analysis of Scan Data}

The analysis of the scan data for {\sscma} largely follows the pattern
we described in Paper~1 for {\syaur}.  The key steps are: (1) identifying
the pixels associated to each trail (star); (2) defining a
minirow-by-minirow detector $ X $ position at each location along the
trail by a one-dimensional fit of the observed signal along each
minirow with a spatially variable line-spread function (LSF); (3) converting the
position into rectified coordinates using the distortion map; (4) removing
the effects of jitter and variable rotation; (5) determining the relative
rectified $ X $ position for each star in each image; (6) combining the
measurements from multiple scans within each epoch, including both
deep and shallow frames; and, finally, (7) estimating the parallax of each
target on the basis of the combined astrometric and spectrophotometric
information.  Key differences in the processing for {\sscma} are: (1) the
availability of an improved geometric distortion solution via the
static correction discussed in \S~2.3.1; (2) the use of serpentine scans,
which are combined with deep and shallow frames for each epoch; and (3)
the introduction of an empirical correction for the $ X $-direction CTE
loss (X-CTE).  Thanks to the number and quality of
reference stars, the nominal error of the parallax for {\sscma} is
significantly smaller than for {\syaur}.  We will now discuss in
detail each step, highlighting the changes with respect to Paper~1.

Our astrometric measurements are based on the spatial scan exposures
listed in Table~3; Figure~\ref{fig:scan_image} shows typical examples of
straight and serpentine scans.

The nearly vertical ``trails'' are the images that each star leaves as
the telescope scans over the field.  The length of the straight scans
is $ \approx 144\arcsec $, 88\% of the length of the field of view of
WFC3/UVIS; thus, the part of the sky covered during the scan is almost
twice the normal field of view of the camera.  Stars near the center
of the region spanned in the detector $ Y $ direction will have trails
that start and end within the frame, while stars farther from the
center along $ Y $ have trails that enter or leave the frame during
the scan.  For serpentine scans, the telescope motion is more complex;
the vertical portions are scanned at $ 1.4 \arcsec \, {\rm s}^{-1} $,
about $ 35\, {\rm pixel} \,\, {\rm s}^{-1} $, a speed chosen to avoid
saturation of {\sscma} in filter \Filter{F606W}.  However, at that scan speed,
and given the desired exposure time of 350~s, the total length of the
scan is $ \approx 525\arcsec $, over three times the field of view of the
detector.  Therefore the scan is folded, consisting of three vertical
legs separated by $ 4\arcsec $ in the detector $ X $ direction.  The
turnarounds for the Cepheid occur just outside the detector field of
view; other stars, at lower (higher) $ Y $ position, can have their first
(second) turnaround within the field of view.

Note that all celestial sources in the field are extended because of
the motion of the telescope.  Cosmic rays are the only compact sources
in the frame and are readily identified by their lack of spatial
extent, allowing us to identify and disregard impacted pixels.

We used a master catalogue of stars to first simulate and then match
the observed trails to these stars.  The fidelity of the simulations
is a few pixels, and in many cases minor adjustments are needed to
ensure that the full 15-pixel window around each trail pixel is fitted.  We
also use the simulations to identify the regions within each star's
trail that are affected by nearby star trails, and disregard the
impacted pixels if the simulation indicates a bias of greater than $ 1.5
\mpix $ for a given row.

The case of serpentine scans is more complex.  For serpentine scans,
each star's trail is marked by multiple reference points: the overall start
and end, which are the start of the first leg and the end of the last
leg, and whose location is determined by the shutter opening and
closing; and the turnarounds between legs, whose location is
determined by the motion of the telescope.  For any given star, only a
subset of these points is visible in the image.  Furthermore, we have
found that the relative location of these reference points cannot be
predicted with sufficient precision from image to image, and therefore
it has to be determined empirically from the data themselves.
Locating individual trails in a serpentine image is thus a two-step
process: first, locate the start, end, and turnaround points for a
subset of well-exposed stars at various locations in the field of
view; second, determine the geometry of the scan (with an accuracy of
$ \sim 3 $ pixels) and apply this geometry to predict the serpentine
trails of all the stars visible in the scan using our star catalogue.
The latter include stars too faint to be profitably measured, but that
are still bright enough to ``spoil'' the scans of brighter stars, as
described above.  The trail map must then be inspected and reference
points tweaked to improve the match with the data; at the end of this
labor-intensive process, we were generally able to identify and locate
the trails of individual stars to within $ \sim 1 $ pixel.

For the purpose of subsequent analysis, trails in serpentine scans are
split into individual passes, each including the trails of all
relevant stars in that pass.  This allows us to associate together
photons collected at the same time, and consistently solve for
time-dependent effects, such as the variable field rotation discussed
below.  It also allows for a more careful identification of
spoilers, since different legs can have different spoiler impact.  In
the case of {\sscma}, each 3-leg serpentine scan results in three
separate position measurements for all the stars, thereby associating
together measurements taken at the same time.  Note that the
serpentine scan was {\it not} obtained during the fifth epoch; thus,
Epochs 1 through 4 each have 7 sets of measured positions (5 for the
Cepheid), while Epoch 5 has 4 sets of positions (2 for the Cepheid).

As in Paper~1, we independently fit each 15-pixel minirow along the
trail to determine the $ X $ position of the star at that value of $ Y
$.  The fit uses an empirical LSF appropriate to the filter and
detector position, obtained by integrating the empirical
point-spread functions (PSFs) from
\citet{bellini11} and, like the latter, oversampled by a factor 4.
Data quality flags from the detector characterization as well as flags
from source-contaminated pixels are used to avoid fitting bad pixels.
The end result is an array of detector $ X $ positions and
uncertainties as a function of the $ Y $-axis position (equivalent to
time) along the scan.

Also following the same procedure as in Paper 1, we start with the
geometric distortion solution for {\Filter{F606W}} from
\citet{bellini11}, which uses a definition of the PSF position that is
consistent with the empirical determination of the PSF itself.  This
geometric distortion map is used to transform the detector $ X $ and $
Y $ positions to sky coordinates.  We obtain a similar solution for
{\Filter{F621M}} from calibration observations of $ \omega $~Cen.  In
addition, in this paper we also correct for the residual geometric
distortion obtained from calibration observations of M67 and M48
(\S~2.2.1).  The original solution from \cite{bellini11} is expected
to have an accuracy of $\sim 0.01$ pixel on scales of $\sim 40 $
pixels; this accuracy is sufficient to reach position precision of 1
{\mpix} ($ 40 \muas $) for full-length scans, which would be a
significant contribution to our overall error budget.  By applying our
new correction, we expect that the local residuals will be reduced to
$ \sim 3 \mpix $ on a scale of 100 pixels, with a projected
contribution to the final error budget of $ \sim 20 \muas $.  Again as
in Paper~1, we use the time-dependent velocity aberration values
provided by the STScI pipeline, interpolated to account for its
variation {\it during} the observation, to correct for the
corresponding plate-scale changes along the scans.  Later, we account
for perturbations in the geometric distortion of the field caused by
the day-night thermal cycle of {\HST} when registering different
scans.

\begin{figure}[h]%t]
\includegraphics[width=\columnwidth,bb=100 62 700 558] {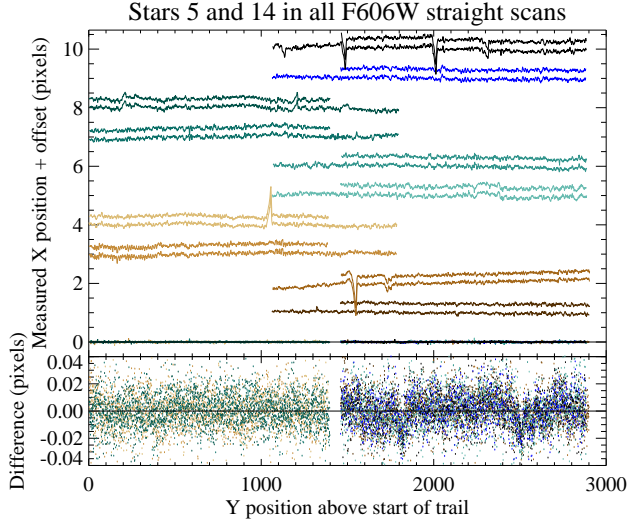}
\caption {\label{fig:separation}
Variations in the measured $ X $ position for two bright stars
(Stars 5 and 14) in the field of {\sscma} in each of the 10 full-length
straight {\Filter{F606W}} scanned exposures.  The top panel shows the individual
X measurements, with a different color for each exposure.  Scans have
been offset along the abscissa to match photons received at the same time,
and along the ordinate by an arbitrary constant.  The very high correlation
between the irregularities in the two scans shows that most of the
apparent ``noise'' in the measured position is actually telescope
jitter.  The differences between the two stars in each exposure are
overplotted at $ X \approx 0 $, and again in the bottom panel with a
scale expanded by a factor of 50.  The pixel-to-pixel variation in the difference is
consistent with the expected uncertainty in the fit for each minirow;
the nominal error in the mean separation in each exposure is 0.4 {\mpix}, or 16
{\muas}.}
\end{figure}

\begin{figure}[h]%t]
\includegraphics[width=\columnwidth,bb=100 62 700 558] {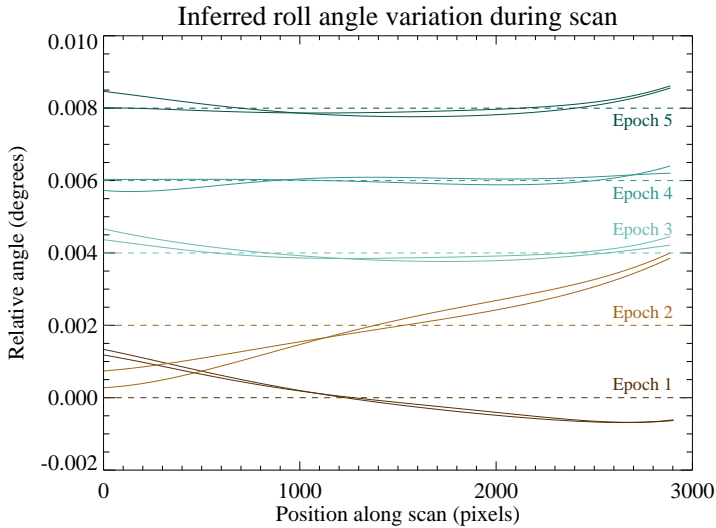}
\caption {\label{fig:dangle}
Differential field rotation during the {\Filter{F606W}} scans
in each of the five epochs of the SS CMa observations, represented as a fifth-degree
polynomial as a function of position along the scan.  A correction
for the differential field rotation is applied to the
measured $ X $ positions along each scan.  Note the similarity of the
pattern for the two observations obtained at each epoch, as well as the marked
decrease in the change starting in Epoch 3, after an improved FGS distortion 
solution was adopted in July 2013.}
\end{figure}

As for the analysis of {\syaur}, we define the one-dimensional
position measurements for each scan line relative to the mean line of
the sample.  This mean line or reference line is determined by
aligning all scan lines in time and taking their weighted average.
The reference line thus contains the jitter history in the direction
perpendicular to the scan which is removed from all lines in their
difference with the reference.  Figure~\ref{fig:separation} shows the
comparison of two bright star trails aligned in scan time and the
residuals after subtraction of one from the other.  We use the
requirement that scan lines, relative to the reference, are parallel
on the sky to measure the time dependence of the scan roll angle.  The
variable rotation history of the scans can be measured well for the
two deep scans obtained at each epoch in {\Filter{F606W}};
Figure~\ref{fig:dangle} shows that the rotation angle is very similar
in both scans obtained at the same epoch.  The variable rotation
cannot be measured as accurately in the shallow scans, because of the
lower signal-to-noise ratio for all stars except for the Cepheid,
which by itself cannot constrain the rotation angle.  As the variable
rotation term appears to be constant within each epoch, we determine
the correction by averaging the measured rotation in the two deep
scans at each epoch, and apply the resulting correction to all deep
and shallow scans for that epoch.  Figure~\ref{fig:dangle} also shows
that the variable rotation is markedly smaller in Epochs 3 through 5,
most likely because of the improved FGS geometric solution adopted on
July~22, 2013, between our Epochs~2 and~3 (Nelan and Lallo, private
communication).

\subsection {Using Multiple Observations at the Same Epoch}

The fundamental measurements at each epoch of observation consist of
the relative $ X $ position of all stars, obtained from the
combination of all coeval scans, deep, shallow, and (if
available) serpentine.  In order to combine these scans, they must be
astrometrically registered, which requires correcting for any
differential geometric distortion.  As in Paper~1, we include in the
distortion model a low-order polynomial correction with free
coefficients for each observation after the first.  We adopt a
polynomial correction to the $X$ coordinate when aligning two frames
which depends on the pixel position of each star trail in the
detector; the assumption is that any variation in the transformation
from true to measured position is tied to the telescope and detector,
and therefore is best described in measured rather than true
coordinates.  Note that a generic first-degree polynomial includes by
definition an $ X $ scale term (the first-order correction in $ X $)
as well as a {\it detector} rotation, which is slightly different but
closely related to the {\it field} rotation previously considered.  As
for M35 and {\syaur}, we find that a second-degree polynomial as a
function of $ X $ and $ Y $ coordinates is adequate to describe the $
X $ coordinate transformation between two scans.  (Only terms of {\it
total} degree up to the polynomial degree are included, so our
second-degree polynomial contains terms in $ X \times Y $, but not in $ X^2
\times Y $ or $ X \times Y^2 $.)  A second-degree polynomial in $ X $ and $ Y $
has five coefficients (plus a constant term), two of which describe an
offset and a rotation, respectively.

For {\syaur} we found that a dearth of intermediate-brightness
stars, combined with the need to allow for second-order polynomial
corrections between the frames, led to significant uncertainties in the
transformation of the Cepheid and of the reference stars to a common
frame, resulting in a dominant contribution to the final uncertainty.

In order to ameliorate this problem, we modified our strategy in two
ways.  First, the number of available intermediate-brightness
reference stars was a primary consideration in the selection of our
targets.  Second, we have obtained and processed serpentine scans in
\Filter{F606W}; these scans provide additional constraints between the Cepheid,
which is not saturated, and a larger number of intermediate-brightness
reference stars.  (Similar data were collected in some of the {\syaur}
epochs, but we were unable to process them properly for Paper~1.)
Position measurements for intermediate and faint stars in serpentine
scans are not quite as accurate as those in the straight {\Filter{F606W}} scans,
for a number of reasons: the same total counts are spread over a
larger area, thus resulting in additional background noise; the higher
density of trails results in more spoilers; and the more complex
telescope motion results in additional parameters to be fitted.
Nonetheless, we find that serpentine scans, while more complex to
analyze, add significantly to the precision and reliability of the
final measurement, and we expect that the analysis of future targets
will use the serpentine scans as well.

To account for a shift in detector $ X $ position due to imperfect
X-CTE, we have obtained calibration observations for WFC3/UVIS in
spatial scan mode, in which a bright star has been moved across the
amplifier boundary at $ X=2048 $ in consecutive exposures.  To the
extent that the X-CTE effect is linear in the distance to the relevant
amplifier, the mean relative position of stars in the field does not
change as the field of view is dithered in the $ X $ direction---if
stars are moved to the right ($ +X $ direction), those to the left of the
boundary ($ X < 2048 $) will experience an increase in their apparent
displacement to the right, and those to the right of the boundary ($ X > 2048
$) will experience a {\it decrease} of their apparent displacement to
the left, for a null net effect.  (Individual stars will move slightly
with respect to one another, as fainter stars will be affected more
than brighter stars.)  However, if a star is moved from left to right
of the boundary, its apparent displacement due to X-CTE will reverse
sign, and thus it will experience a large net motion.  An analysis of
calibration observations obtained in Fall 2014 shows that 
the net effect is $ 8 \mpix $ for a star near
saturation, thus implying that a bright star near the amplifier
boundary was shifted by $ 4 \mpix $ at that time owing to X-CTE.

Experience with WFPC2, ACS, and (more limited) with WFC3/UVIS
strongly suggests that CTE effects grow linearly with time in orbit
and with distance to the relevant amplifier.  For WFC3/UVIS, we assume that
the CTE loss was zero at launch.  If this is correct, and
if the effect is furthermore anti-symmetric with respect to the
amplifier boundary, the impact of X-CTE upon parallax determinations
vanishes {\it as long as the observations are obtained with the same
center and rotated field of view}.  The reason is that a growing X-CTE
will result in an apparent motion for each star which is
indistinguishable from a proper motion; thus, each star will have a
spurious term in its estimated proper motion, but the parallax
estimate is unaffected.  Even if the field of view is shifted, the
impact on each star is minimal, as discussed above---as long as the
star does not switch amplifiers.

However, second epoch observations placed
the Cepheid at the same $ X $ detector location ($ X \approx 2000 $, left of
the amplifier gap) as the odd epochs, thus shifting the field center 
by about 100 pixels between the two orientations.  The aim was to
minimize the impact of uncertainties in the geometric distortion for
the Cepheid by placing it in approximately the same detector location.
The magnitude of the X-CTE effect was not fully understood at that
time.  On the basis of current information, minimizing the impact of
X-CTE effects is deemed more important, and starting from Epoch~3,
observations were obtained with the same field center for all epochs.

In order to correct for the residual X-CTE effect---which generally
only affects the Cepheid target, as others stars within 50 pixels of
the amplifier boundary are swamped by the light of the target---we
simply correct the position measured for the Cepheid in Epoch~2 by
twice the estimated offset at that time, about $ 2.9 \mpix $.  This
correction mimics the effect of placing the Cepheid in the symmetrical
position ($ X \approx 2100 $) and thus nullifies the effect of X-CTE
on the measured parallax.  As discussed above, relative proper motions
{\it will} be affected by X-CTE and thus can only be determined
accurately if a good overall calibration for X-CTE is obtained.

\subsubsection {Serpentine Scans}

Although in many ways the serpentine scans are treated similarly to
the regular straight scans, some special considerations apply.  For
each leg, we exclude from the fit a region of about 300 pixels before
and after each turning point, in which the motion of the telescope
deviates significantly from a straight line.  In principle, we could
include this deviation in the overall fit; however, the local slope
can be large enough that our underlying approximation that the scan
direction is perpendicular to the resolution direction no longer fully
applies.  For such regions, the variable-rotation solution (see \S~2.4)
would also fail its underlying assumptions.  Therefore, we simply
ensure that we only include the portion of each leg where the mean
displacement of the motion from a straight line is less than 0.1
pixels.

In addition, serpentine scans do not provide a good way to determine
the start or end point of each leg accurately during the initial fit.
The half-rise method does not work, as each leg is truncated by the
cutoff in the horizontal offset, rather than by the rise or fall due
to the shutter.  Only the very first and last leg could have a
half-rise measurement, and only if the relevant start/end point occurs
on chip.  For this reason, we put special care in estimating the start
and end point from the overall shape of the serpentine scan, and we
use these start and end points for the initial guess at the vertical
positioning of each scan.  In keeping with our procedures for straight
scans, a refinement step occurs in which a least-squares fitting of
the jitter pattern along the scan is used to improve the relative
positioning of each scan.  This step works to a similar accuracy as
for the straight scans, after taking into account the relative signal
levels and scan lengths.

\subsection {Combining Multi-Epoch Data: Toward a Parallax Measurement}

For each epoch of observation, our goal is to obtain a measurement of
the relative positions of all stars, the Cepheid as well as all the
reference stars, along the resolution direction (the
distortion-corrected detector $ X $ axis projected onto the sky).
This measurement must be as accurate and free of systematics as
possible.

In order to obtain a measurement of the relative parallax and proper
motion (in the resolution direction) of the stars on the field, data
from all available epochs must be combined.  In addition, information
on the distance of the reference stars is required in order to obtain
an absolute parallax for the target.  In principle, the process
requires only a linear combination of the positions as measured at
each epoch to determine relative parallaxes and proper motions; if the
mean parallax of the reference stars can be estimated, determining the
parallax of the target is then straightforward.

In practice, solving for the parallax of the target is much more
complex.  Small rotations (at the level of a few hundredths of a
degree) between epochs, as well as small changes in plate scale and
low-order geometric distortion which are known to occur, can
substantially affect the projected positions of each star in each
epoch, and thus impact significantly its estimates of parallax and
proper motion.  Occasionally, reference stars can have anomalous data,
either because of measurement problems (e.g., undetected faint stars
close enough to affect the fit), or for astrophysical reasons (e.g.,
binary companions or other sources of photocenter motion).

The approach we have adopted solves {\it simultaneously} for the
astrometric parameters of all the stars in the field and for the
geometric registration of all the epochs, using the spectrophotometric
parallax estimates as priors for their astrometric parallax.  With
this approach, the spectrophotometric parallax estimates help
constrain the registration between epochs, including the relative
low-order geometric distortions.  It is thus critical that the
parallax estimates be as accurate and robust as possible.

In the next section, we present the combination of spectroscopic and
photometric data and the analysis that leads to spectrophotometric
distance estimates for as many of the reference stars as possible.  

In \S~4 we return to the determination of the Cepheid parallax using
the spectrophotometric distance estimates obtained in \S~3.

\section {Spectrophotometric Data and Distance Estimates}

\subsection {Photometry and Spectroscopy of Reference Stars}

Narrow-angle astrometry, such as what we can obtain with {\HST}, is
fundamentally differential in nature, and therefore it can only
constrain the {\it difference} between the parallax of stars within
the field of interest.  In order to convert this relative parallax
estimate into an absolute measurement, the parallax of other stars in
the field must be estimated, and careful consideration must be given
to possible systematics and random uncertainties in these estimates.
In addition, estimates of the parallaxes of other stars in the field
can {\it confirm} the quality of the astrometric measurements,
identify outliers, and help constrain some of the low-order geometric
distortion variations discussed earlier.

For this reason, we have obtained multi-band photometry and
medium-resolution, classification-quality spectroscopy for most likely
reference stars in the field of SS CMa.  A combination of stellar
model fitting and spectrophotometric classification, together with an
understanding of the distribution of stars along the line of sight,
has been used to estimate the distance to each reference star and its
likely uncertainty.  We also used prior estimates of the reddening
along the line of sight, based on measurements of stars in 2MASS and
Pan-STARRS, in order to constrain the range of possible reddening;
however, the final reddening-distance law was also fitted for in our
analysis.

\subsubsection {Photometry for Reference Stars}

For the {\sscma} field (and for other Cepheid fields in progress), we
obtained direct imaging with {\it {\HST}} during the scanning
observations and measured photometry of all reference stars in the UV
({\Filter{F275W}}, {\Filter{F336W}}), Str\"{o}mgren
({\Filter{F410M}}, {\Filter{F467M}}, {\Filter{F547M}}), and broad-band
({\Filter{F850LP}}) systems.  In order to obtain the photometry
efficiently within the observing time available to our program, we
used $ 2 \times 2 $ binned mode, in which full-field WFC3/UVIS images are binned
on-board before being saved to the {\HST} computer for download.  In this
mode, images have a substantially smaller memory footprint, and more
images can be obtained before the instrument memory is full and the
images must be transferred to the {\HST} solid-state storage.  
Consequently, we were able to obtain several photometric measurements within
each orbit, without impacting the scanning mode observations.  We have
developed and tested procedures to accurately recover and calibrate
binned-mode photometry, and we obtained reliable photometry (albeit
with larger uncertainties) even for partially saturated stars.

We also obtained {\Filter{F160W}} photometry with WFC3/IR, and we
added $ J, H, $ and $ K $-band photometry from the 2MASS survey,
as well as Channel 1 and Channel 2 photometry
from {\it WISE} when available, to provide a set of up to 14 bands of
photometry from $ 0.2 $ to $ 4.5\, \micron $.  All of the photometry was of
high signal-to-noise ratio, with the exception of {\Filter{F275W}} where only a third of
the stars yielded a measurement ($ {\Filter{F275W}} < 22.8 $ mag).
Missing or excluded photometry was recorded for stars which suffered
cosmic ray hits, suffered blending in the 2MASS or {\it WISE} data (as
identified from their data quality flags or from 
{\it {\HST}} {\Filter{F850LP}} imaging), and for half the
field not covered by {\Filter{F410M}} imaging.  The resulting
photometric information is reported in Table~2.

\subsubsection {Spectroscopy of Reference Stars}

We independently determined the temperature and luminosity class of
the majority of the reference stars via medium-resolution optical
spectra compared to template spectra.  As indicated in Table~2, 
spectra were obtained with the Kast double spectrograph \citep{miller93}
on the 3~m Shane reflector at Lick Observatory, with GMOS on Gemini South
\citep{hook04}, and with LRIS on Keck \citet{mccarthy98}.
Standard procedures were used for the data reduction.

\subsubsection {Estimating Spectrophotometric Parallaxes}

Spectroscopic parallaxes of stars in the field were determined, as in
Paper~1, by matching up to 14 bands of photometry to stellar
isochrones, comparing medium-resolution spectroscopy to stellar
spectra for classification standards, and using the Besan{\c{c}}on
Galaxy Model \citep [and references therein]{robin86, robin03} as a
likelihood prior for stellar parameters.  We used a version of the
model with an updated thick disk which better fits Sloan Digital Sky
Survey (SDSS) and 2MASS data (Robin 2013, private communication).

\begin{figure}[ht]
\includegraphics[width=\columnwidth,bb=100 62 700 558] {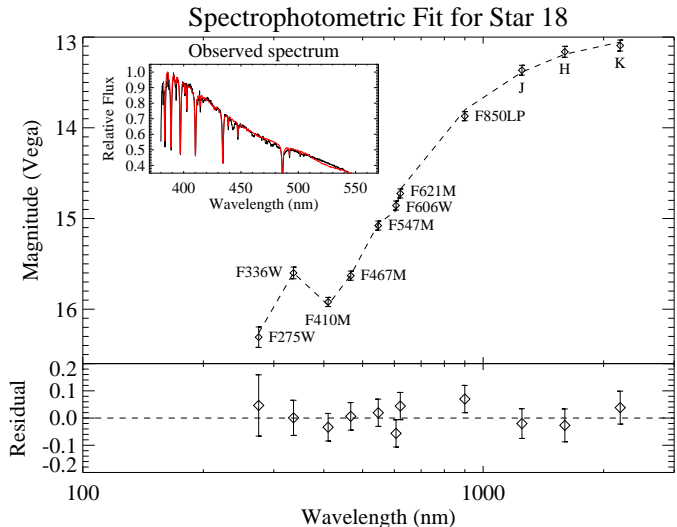}
\caption {\label{fig:spectral_fit}
Spectral and photometric fit for the B2~V star~18.  (Top panel) Observed photometry
(diamonds) and best-fitting model from the Padova isochrones (dashed line).
(Inset) The observed spectrum (black), continuum-corrected and fitted to
a model spectrum (red) using MKCLASS.  (Bottom panel) Residuals of the
model photometric fit, shown on a larger vertical scale.  Photometric errors
include a 0.05~mag term added in quadrature to the measurement uncertainty
to account for possible differences in photometric systems
between models and observations.}
\end{figure}

Our procedure for measuring the spectroscopic parallaxes of the
astrometric reference stars in the field has been somewhat refined and
improved since the procedure used for the field of SY Aurigae
described in Paper~1.  We have now added photometry of the stars from
two bands of spatial scanning, a broad (\Filter{F606W}) and another
Str{\"o}mgren (\Filter{F621M}) filter, one broad band from {\HST} in
the NIR (\Filter{F160W}), and two bands of medium-IR data from the all
sky {\it WISE} mission.  We add an uncertainty of 0.05~mag in quadrature
to all photometric uncertainties
to account for possible differences in photometric systems
between models and observations.  The stellar classification of star
temperature and luminosity class is now done using the MKCLASS version
1.7 automated Morgan-Keenan classifier \citep{gray14}.  Finally, we
have improved our prior knowledge of the extinction along the line of
sight as a function of distance using the 2MASS determinations from
\citet{marshall05}, who provided extinction vs.~distance estimates for
the line of sight in the direction of SS CMA (private communication)
with an uncertainty of 0.3 mag at a given distance.  An example of the
quality of the results is shown in Figure~\ref{fig:spectral_fit} for
Star~18; the observed photometry (diamonds) is matched to a reddened
model (dashed line), with the residuals shown on a larger scale in the
bottom panel.  The inset shows the observed spectrum for Star~18
black) overlaid with the best-fitting model according to MKCLASS
(red).  Note that the

\begin{figure}[ht]
\includegraphics[width=\columnwidth,bb=100 62 700 558] {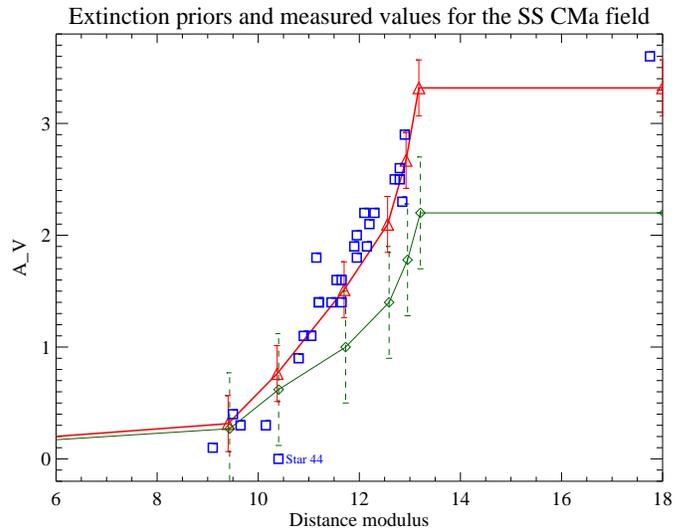}
\caption {\label{fig:av_dmod}
Estimated extinction vs. distance modulus for the {\sscma}
field.  The green curve and values (diamonds with error bars) 
show the starting estimate of the relation between extinction and
distance, which is iteratively adjusted during the fit process.
The red curve and values (open triangles with error bars) shows
the final relation.  The blue squares show the estimated extinction for
each of the reference stars with sufficient spectrophotometric information.}
\end{figure}

Because it can be difficult to estimate extinction along a line of
sight at very low Galactic latitude, and because our up-to-14 band
photometry stellar photometry spanning the 0.275 to $ 4.5 \,\micron $
can aid the determination, we started with a weaker prior having a
Gaussian width of 0.5 mag, and then, on the basis of the {\it a
posteriori} extinction estimates, we applied a global correction to
the 2MASS estimates before reverting to the 0.3 mag uncertainty for a
final estimate.  The maximum extinction we allowed in the fits was 1.2
times the total extinction to infinity along this line of sight
estimated by \citet{schlafly11}, which in turn was based on a
rescaling of the IRAS-based estimate of \citet{finkbeiner98}.  The
scatter of the {\it a posteriori} extinction estimates for each star
around the extinction prior at its (spectrophotometrically) estimated
distance was 0.21~mag, with extinction values ranging from $ A_V
\approx 0 $ at distance modulus $ \mu = 9 $ to $ A_V = 3 \,\rm mag $
at $ \mu=13 $.  Figure~\ref{fig:av_dmod} shows the resulting final law
for $ A_V $ vs.~$\mu $ (red points and red line), together with the
individual values for each of the reference stars in the field (blue
squares).  The relation thus obtained between extinction and distance,
for stars both closer and further away than the Cepheid, will also
serve to constrain the reddening estimated for the Cepheid.  This will
in turn provide information on the intrinsic colors of the Cepheids
and improve the robustness of the global {\PL} calibration we expect
to obtain.

\section {The Absolute Parallax of {\sscma}}   % Section 4

\subsection {Multi-Epoch Combination and Parallax Fit}

The final step in the astrometric solution consists of combining the
multiple measurement epochs taken over the course of two years at
intervals of six months to fit to our standard astrometric model which
involves three parameters for each star: position, parallax, and
proper motion along the measurement direction.  The results we present
here for {\sscma} are based on five epochs of observation; four more epochs
are being obtained as part of a recently approved program extension
(Program GO-14206).  The fifth epoch does not include the serpentine scan.
The exposures used are listed in Table~3.

Together with the astrometric parameters of each star, the model
includes up to second-order geometric parameters used to align each epoch with one
another (offset and rotation), as well as any residual large-scale
adjustment to the geometric distortion required to reduce the model
residuals.  This last part is identical to the single-epoch aggregation 
step, but it now substitutes the stationary-star assumption
with the astrometric model for each star.  

The full model can be formally described by the expression
\beqa
X_{ij} = X_{i0} - X_{{\rm ref},j} + PMx_i \, (t_j-t_0) + \\ 
\null \quad \quad \pi_i \, f_j + R_j\, Y_{i0} + \langle P_j (X_{\rm det}, Y_{\rm
det}) \rangle_{{\rm trail}_{i}} \nonumber
\eeqa
\noindent
where the basic measurements are the positions $ X_{ij} $---that is,
the $X$ position of the trail of star $ i $ in image $ j $ (relative
to the reference scan line), measured after correction for variable
rotation, scale-corrected for velocity aberration and variable
distortion, and projected onto a constant sky frame.  The $ X $
coordinate is aligned with detector $ X $ and, by design, aligned with
the bulk of the parallactic motion.  The quantity $ X_{i0} $ is the
reference position of star $ i $ at time $ t_0 $, and $ X_{{\rm
ref},j} $ is the offset of image $ j $ in the $ X $ direction---in
essence, the position of the reference scan line for image $ j $ on
the sky.  The astrometric motion of star $ i $ in the $X$ direction is
described by the $ X $ component of the proper motion, $ PMx_i $, and
the parallax $ \pi_i $, applied with the epoch-dependent parallax
factor $ f_j $.  The term $f_j$ is the projection (for unit parallax)
of the parallactic motion in the $X$ direction at the time of the
observations, computed using the formulae at pp.~B28 and C5 of {\it
The Astronomical Almanac} (2013), and the orientation of the detector
axes from the image headers.

As we did for \syaur, the model position must be corrected for the
relative rotation and geometric distortion of image $ j $ with respect
to the reference image.  The rotation term on the sky is $ R_j \,
Y_{i0} $, where $ R_j $ is the rotation of image $ j $ and $ Y_{i0} $
is the static relative position of star $ i $ in rectified coordinates
along the $ Y $ direction with respect to the center of the field.  We
find our rotations to be of order $ 10^{-5} $, so even a coarse
measurement of $ Y_{i0} $ with a precision of $ \sim 1 $~pixel will
suffice.  The polynomial term is determined simultaneously with the
astrometric parameters during the model-fitting procedure, as a
second-degree $ P_j (X_{\rm det} , Y_{\rm det} ) $, where $ X_{\rm det}
$ and $ Y_{\rm det} $ are detector coordinates; the total correction
is determined by evaluating the polynomial for image $ j $ at every location
along the trail of star $ i $ in that image, and averaging the result.
The constant term is omitted from the polynomial because it is degenerate with 
the image offset $ X_{{\rm ref},j} $.

Our proper motion term is relative to the set of stars in the field
and contains a contribution from the estimated X-CTE term per year at
that star's location.

Note also that the model is formulated to be {\it linear} in the
astrometric parameters, which in turn are linearly related to most
measured quantities, i.e., positions on the detector.  As a
consequence, the errors in the derived astrometric parameters for the
Cepheid are likely to be very nearly Gaussian (the same does not
necessarily apply to distant stars for which the spectrophotometric
constraints dominate the error distribution).  As long as the parallax
and its error distribution are used directly, there is no need to
apply nonlinear correction such as those suggested by \citet{lutz73}.
More generally, proper consideration of all prior information used in
selecting and characterizing the population of our target Cepheids
will be required in determining the optimal calibration for the {\PL}
relation on the basis of our measurements (\citealt {hanson79,
francis13}; see also the discussion in \citealt{benedict07}).
Therefore we do not include a Lutz-Kelker-type correction for the
parallax of {\sscma} at this point, but defer consideration of the
proper characterization of the prior probability density function
for our target to the analysis of the full sample of Cepheids.

As far as the astrometric model is concerned, parallaxes are also
relative; however, the degeneracy in the conversion to absolute
parallaxes can be broken by using the spectrophotometric distance
estimates for the stars in the field discussed in \S~3.  The
distance estimate of the target star will be insensitive to
uncertainties in the distance of the reference stars so long as the
set contains objects which are bright and distant (e.g., red giants).
Note also that in addition to providing a conversion to absolute
parallax, individual spectrophotometric parallaxes are also helpful in
constraining some epoch-to-epoch geometric transformations.  We will
discuss in detail in \S~4.2 the impact of the parallax
constraints for reference stars on the multi-epoch solution, and
investigate the consequences of uncertainties, outliers, and other
possible issues.

Each epoch after the first is allowed a rotation, a net offset,
and a second-degree polynomial adjustment to match the first
epoch; since there are about 20 stars useful for measurement at each
epoch, these additional 7 parameters per epoch over which we
marginalize do not place an undue burden on the solution.

Formally, the {\it a priori} distance estimates based on
spectrophotometric parallaxes serve as Bayesian priors for the
parallax of the stars in the field.  A prior is not used for the
Cepheid, so that its distance estimate is determined directly and only
from its observed parallax.

The best values of the model parameters are determined by minimizing
the total model $\chi^2$ to achieve the most likely parameters.  Among
these parameters is the {\it absolute} parallax of {\sscma}, which results
directly from the model optimization.  A
modest fraction of the reference stars in the field are expected to be
part of binaries with parameters that would cause a significant
deviation from our simple astrometric model.  This fraction depends on
distance and spectral class, but is $ \sim 10$--20\% for F and G stars
at $ 1 \kpc $ on the basis of the distribution of binary properties in
\citet[see also discussion in Paper~1]{duquennoy91}.  We run the global
model iteratively after rejecting one outlier (Star~n2) on the basis
of its disproportionate contribution to the total $ \chi^2 $.  We also
exclude Star~n10 because its astrometric parallax is suspect,
resulting in a very large estimated distance which is inconsistent
with its spectrophotometric information.

\begin{figure*}[ht]
\includegraphics[width=\textwidth, bb=0 0 830 558] {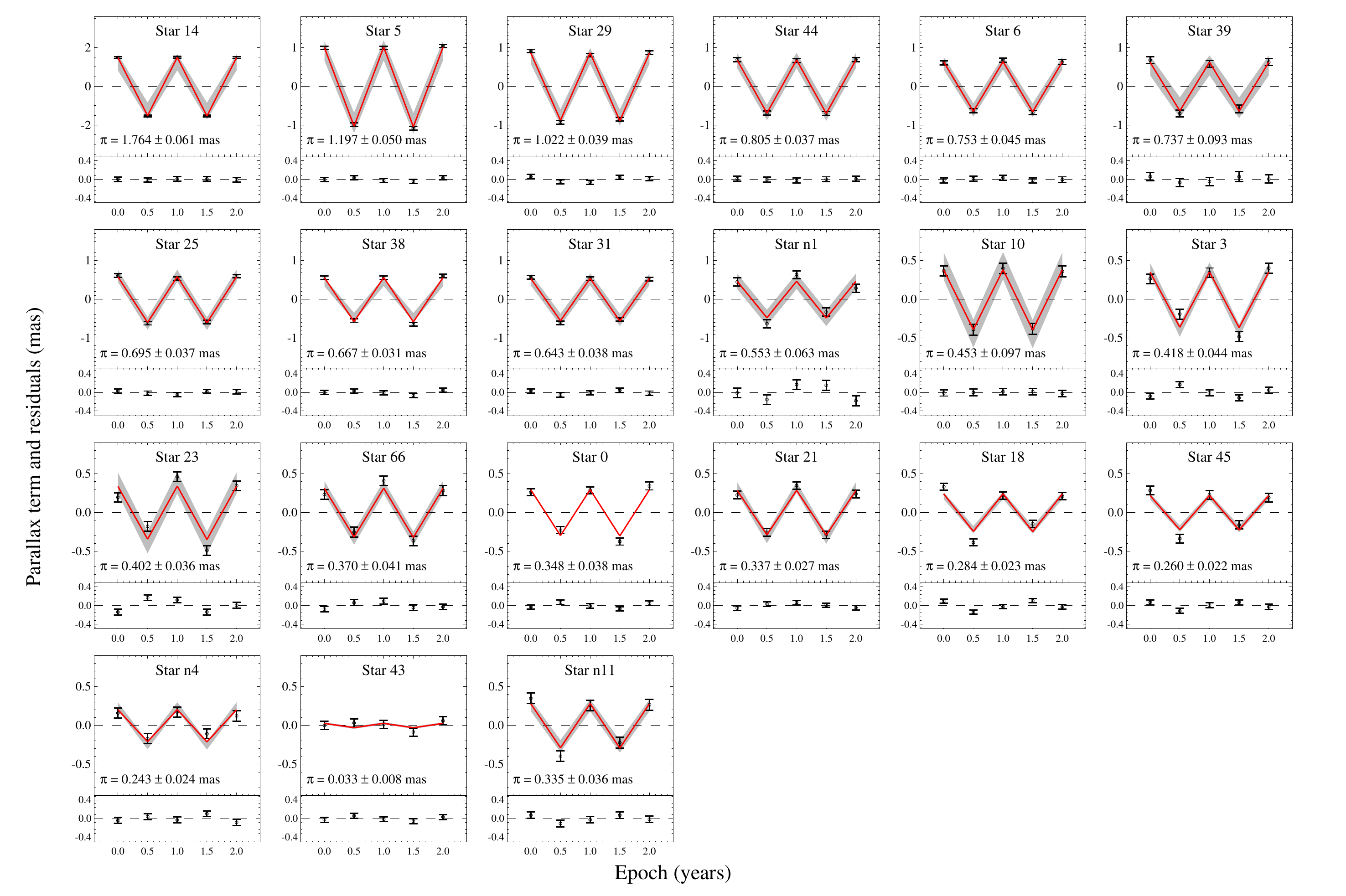}
\caption {\label{fig:wfig}
Individual stellar parallaxes in the field of {\sscma}.
The red line indicates the measured parallax; the grey band 
indicates the spectrophotometric parallax with $\pm 2\sigma$
width.  The Cepheid {\sscma} is Star~0.  Fitted proper
motions have been subtracted from the measurements
and fits for ease of viewing.  Star~29, the putative companion of the
Cepheid, is much closer to the Sun and is not physically associated with it.}
\end{figure*}

\subsubsection {Multi-Parametric Model for {\sscma}: Primary Solution}

Figure~\ref{fig:wfig} shows the best estimate of the parallax for the
stars in the {\sscma} field.  For reference stars, the reported
parallax combines both astrometric and spectrophotometric information;
no spectrophotometric information is used for the Cepheid {\sscma}
(the star labeled 0).

For each star, the top panel shows the astrometric measurements at
each epoch (dots with error bars) and the best-fitting parallax model
(red line), both in milliarcseconds; the fitted proper motion is subtracted
from both measurements and model for ease of display.  The gray band
shows the spectrophotometric parallax estimate with a $ 2\,\sigma $ uncertainty,
when available.  The bottom panel shows the astrometric residuals from
the best model, also in milliarcseconds.

The best estimate of the parallax of {\sscma} is $ 0.348 \pm 0.038
\mas $, corresponding to a distance estimate of $ 2.87 \pm 0.33 $ kpc.
However, note that, as commented in \S~4.1, the error distribution is likely
Gaussian {\it only in parallax}.  A nonlinear conversion, e.g., to distance,
will have a nonsymmetric error distribution, which must be taken into
account in further processing.  It is also necessary
to consider any prior information used in selecting and characterizing
the sample \citep{hanson79, francis13}; see also the analysis in \citep{benedict07}.

\begin{figure}[ht]
\includegraphics[width=\columnwidth,bb=100 62 700 558] {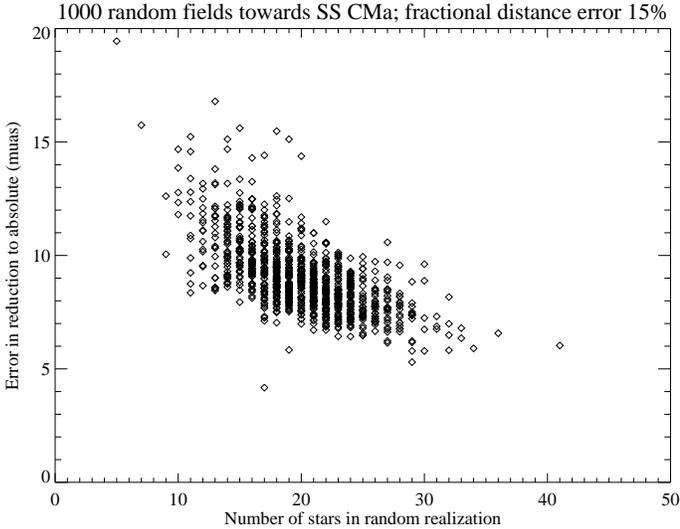}
\caption {\label{fig:conversion_to_absolute} Distribution of expected
errors in the conversion to absolute parallax for 1000 fields drawn randomly
from the Besan\c{c}on model in the direction of {\sscma}.  We assume
that only half the stars in each field are available as reference
stars, and that each has a 15\% error in its spectrophotometric
distance estimate (about 0.3~mag in the estimated absolute magnitude).
Typical values are $ 9 \muas $ for $ \sim 20 $ stars, consistent with
the actual values for our field when Star 43 is excluded.}
\end{figure}

The uncertainty in the conversion to absolute parallax (i.e., the
systematic uncertainty in the frame parallax) of the set of 20 fitted
reference stars is $ 7 \muas $, well below our target uncertainty.  In
the field of {\sscma}, the precision of the conversion to absolute
parallax benefits from the presence of some very distant stars, and especially
Star~43, whose spectrophotometry indicates that it is a K giant at
around 30 kpc (\S~4.3).  However, even excluding Star~43, the rest of
the reference stars indicate an uncertainty in the frame parallax of
about $ 11 \muas $, still much smaller than our target uncertainty.
This is not surprising; Fig~\ref{fig:conversion_to_absolute} shows the
typical precision of the correction to absolute parallax in random
fields generated from the Besan\c{c}on model for the direction of
{\sscma}, assuming a 15\% typical uncertainty in spectrophotometric
distance estimates and that only half of the stars in the field will
be available for the conversion.  A typical field would have $\sim 20
$ available reference stars and an uncertainty of $ 9 \muas $ in the
conversion to absolute parallax for an assumed RMS uncertainty of 0.3 mag
in the estimated distance moduli, comparable to the values for the
actual data.  (The uncertainty in the conversion to absolute scales roughly 
linearly with the assumed uncertainty in the distance moduli.)
However, Star~43 is unusual in its own right, and will
be further discussed in \S~4.3.

\subsection {Astrometric vs.~Spectrophotometric
Parallax; Partially-Constrained Solutions}

In the primary solution, we assumed that the spectrophotometric and
astrometric distance estimates for each star are separate but valid
measurements of the same quantity, the physical distance of each star.
In practice, this assumption may not always be true, e.g., because of the
possible binarity discussed earlier.  On the other hand, the assumptions
underlying our spectrophotometric distance estimates may not be valid
for all stars, and there could be outliers---e.g., due to anomalous
extinction, or a history of mass exchange---which could occasionally
lead to a faulty estimate of the distance modulus.

A careful comparison of astrometric and spectrophotometric parallaxes
for the reference stars can provide a powerful check of our procedures
and our final accuracy.  Spectrophotometric parallaxes, derived from a
combination of spectra and multiband photometry in conjunction with
stellar model tracks, a model of the density distribution of stars
along the line of sight, and an extinction model, typically have
fractional accuracy that varies little as a function of distance; thus,
their absolute error is much smaller for distant stars than for nearby
ones.  On the other hand, astrometric parallaxes have absolute errors
that are similar in magnitude in terms of parallax angle, and hence
nearly independent of distance, although they do scale with
apparent brightness; thus, for nearby stars, astrometric parallaxes are
more accurate than spectrophotometric ones, and vice versa for distant
stars.  For typical stars in our analysis, the two accuracies are
comparable at about 1 kpc.  Consequently, stars beyond 1 kpc provide a
solid check of astrometric measurements, while closer stars provide a
verification of the spectrophotometric estimates.

However, the parallax estimated for each star in the full (primary)
solution shown in Figure~\ref{fig:wfig} is affected by its
spectrophotometric prior, and hence cannot be used directly for an
independent check of the astrometric parallax thus obtained.  On the
other hand, dropping all of the spectrophotometric priors is not a
viable option.  First, all narrow-field parallax measurements are, by
necessity, only relative; thus, without {\it some} spectrophotometric
measurements, no absolute parallax can be derived.  Second, as
discussed in \S~4.1, without a spectrophotometric prior for the
majority of the stars, we lack the ability to properly constrain the
relative alignment and polynomial distortion for each epoch of
observation, thus worsening the quality of the measurements and
introducing substantial degeneracies in the solution process.

In order to carry out a meaningful test of the quality of our
astrometric parallaxes, we repeat the multi-epoch fit by discarding
the spectrophotometric prior for each star in turn, and define the
parallax obtained for that star as its {\it pure astrometric}
parallax.  For example, when measuring the pure astrometric parallax
for Star~29, we discard the spectrophotometric prior {\it only for
Star~29}, but retain the prior for all the other stars in the field
that have one.  This allows the solution to converge with only a minor
decrease in overall precision, resulting in a trigonometric parallax
estimate to Star~29 that is completely independent of any photometric
or spectroscopic information for that star.  We call this the ``pure''
astrometric parallax, and list it as $\pi_{\rm astro} $ in Table 2,
where the parallax resulting from the full solution is labeled $
\pi_{\rm full} $.  We then repeat this process for all other stars for
which a spectrophotometric prior is available; for stars without a
spectrophotometric prior, including the Cepheid, the pure astrometric
parallax is of course identical to the parallax from the full
solution.  This procedure not only allows us to assess the quality of
our astrometric measurements with the spectrophotometric distance
estimates, but also mimics the handling of the Cepheid itself, for
which no spectrophotometric prior is ever used, and thus provides a
useful test of the validity of its parallax measurement.

\begin{figure}[ht]
\includegraphics[width=\columnwidth, bb=90 62 570 558] {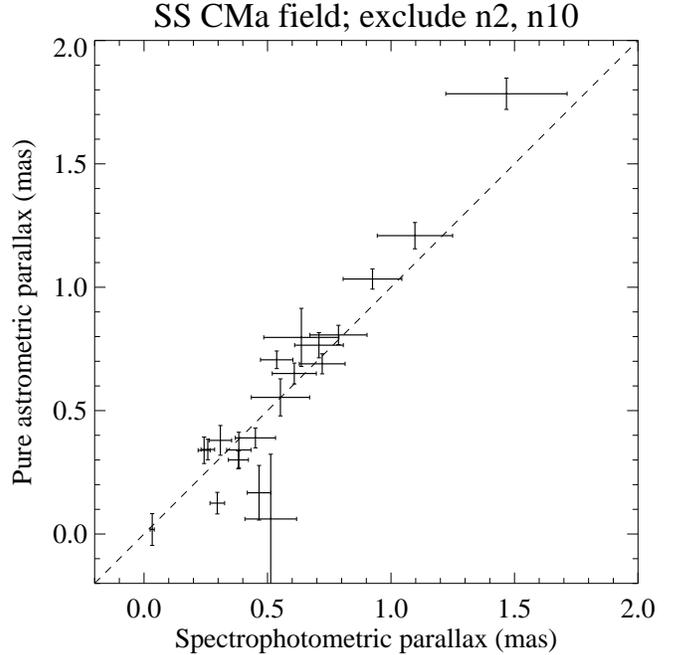}
\caption {\label{fig:pxdiff} Comparison between spectrophotometric and
pure astrometric parallax, obtained by excluding the
spectrophotometric prior for each star in turn.  Therefore, the
pure astrometric parallax for each star is solely based on its
astrometric measurements, and is not affected by any spectroscopic or
photometric data for that star.}
\end{figure}

Figure~\ref{fig:pxdiff} shows a comparison of the pure astrometric and
spectrophotometric parallaxes for the reference stars in the field of
{\sscma}.  The Cepheid is not included, as no spectrophotometric prior
is used for it.  For the majority of the stars, there is a very
reasonable agreement between them, with 17 out of 20 within nominal
$2\,\sigma$; only stars 3, 38, and n4 are outside this range, and
Star~10 has a very small value for the pure astrometric parallax,
with a very large uncertainty.  

Excluded from Figure~\ref{fig:pxdiff} is Star~n10, for which the pure
astrometric parallax is negative.  In our solution, ``negative''
parallaxes are not necessarily disallowed; they can occur, for
example, if the correction to absolute parallax is underestimated, and
all stars are in reality closer than the astrometric parallax
indicates. In that case, the true parallax would simply be larger than
the resulting value, because of the larger correction to absolute
parallax.  However, in this case the solid agreement between
astrometric and spectrophotometric parallaxes for most stars argues
strongly against a large systematic error in the reduction to absolute
parallax.  Although we could limit the solution to require {\it
positive} parallaxes, this step would be somewhat arbitrary, given
that the actual value of the reference parallax is part of the
optimization process, and could give excessive weight to stars with
very low parallaxes in the solution.  Therefore we treat Star~n10 as
an astrometric outlier and exclude it from our solution.  Based on the
discussion in Paper~1, we expect $\sim 10\% $ of the reference stars
to be astrometric outliers due to binarity, so the presence of an
outlier in this sample is not surprising.

\begin{figure}[ht]
% \vspace*{180mm}
\includegraphics[width=\columnwidth,bb=100 62 700 558] {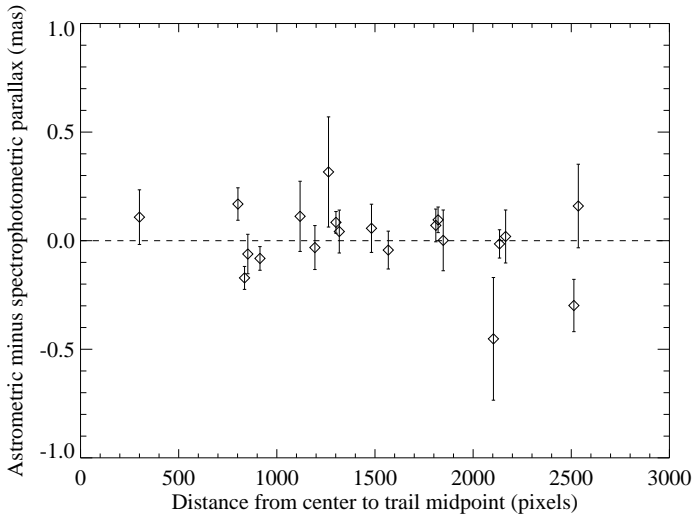}
\caption {\label{fig:pxdiff_dcenter} Difference between
spectrophotometric and pure astrometric parallax, plotted as a
function of the distance of each star from the field center.  The
error bars combine in quadrature the uncertainties in both parallaxes.  
The largest deviations occur for stars far from the center of the field,
reflecting the difficulty in constraining the polynomial distortion
terms when the stars at the edge of the field (which have the most
leverage) are excluded from the prior.}
\end{figure}

It is important to note that the accuracy of the partially constrained
solutions obtained by dropping the spectrophotometric prior for one of
the reference stars can be potentially compromised, especially if that
star is near a corner of the field.  The reason is that the low-order
polynomial distortion that we adopt to register the measurements
across epochs may lack a critical constraint near that corner, while
on the other hand its {\it value} is required at that location.
Consequently, a quasi-degeneracy in the multi-epoch astrometric
solution exists for that star.  This is especially apparent for stars
such as Star~3, which is the only bright star near the bottom left of
the field (see Fig.~\ref{fig:direct_image}).  The full solution
(Fig.~\ref{fig:wfig}) shows for Star~3 a normal parallax value and
uncertainty of $ 0.418 \pm 0.044 \mas $, and its residuals are fairly
typical.  On the other hand, the partially constrained solution in
which its parallax prior is dropped is $ 0.167 \pm 0.111 \mas $, with
a very different value and a much larger uncertainty than the fully
constrained solution.  Inspection of the two solutions shows that they
differ by about 40\% in the $ Y^2 $ polynomial term, resulting in a
differential offset of up to $ 10 \mpix $ in Epoch 5.  The case of
Star~10 is even more extreme, with an increase of the nominal error
from $ 0.096 \mas $ to $ 0.260 \mas $ for the pure astrometric
parallax.  For most other stars, the uncertainty increases by 10--70\%
when the spectroscopic prior is dropped.  (Another exception is
Star~43, which has a comparatively poor fractional accuracy in the
trigonometric parallax because of its very large distance.)
Figure~\ref{fig:pxdiff_dcenter} shows the difference between pure
astrometric and spectrophotometric parallax as a function of distance
from the center of the field: two of the three stars with difference
larger than $ 0.3 \mas $ are more than 2000 pixels from the field
center.

In summary, we conclude that there is in general good consistency
between the astrometric parallaxes we measure from this set of {\it
HST} spatial scans and the distance estimates obtained from our
spectroscopic and multiband photometric measurements.  Moreover, as
expected, the availability of such estimates for a large fraction of
the reference stars is necessary to constrain the overall astrometric
solution.

\subsection {Two Special Reference Stars}

Among the reference stars in the field, Star~29 was suggested by
\citet {evans94} as a possible binary companion to {\sscma}, on the
basis of its blue color, estimated distance, and likelihood of the
relatively small separation of $ 13 \arcsec $.  However, we classify
Star~29 as G6~V on the basis of its spectrum (see Table~2), and its
estimated spectrophotometric distance modulus is $ 10.2\pm 0.3 $ mag,
which places it significantly closer than the Cepheid.  As shown in
Figure~\ref{fig:pxdiff_dcenter}, the astrometric parallax is in
agreement, and we conclude that Star~29 is {\bf not} physically
associated with the Cepheid {\sscma}.

Another interesting reference star is Star~43.  Although faint in the
visible ($ V_{606} = 15.26 \, \rm mag $), this star is quite bright in
the NIR ($ H = 10.76 \, \rm mag $), and is classified from the
spectrum as a K5~III, with spectrophotometric distance modulus $ 17.4
\pm 0.5 $ mag.  This places the star about 30 {\kpc} from the Sun,
well outside the disk and the spheroid of the Galaxy.  (Note that this
star by itself carries about half the weight of the conversion to
absolute parallax for this field.)  The trigonometric parallax without
spectroscopic prior is $ 0.018 \pm 0.065 \mas $, hence not
significantly detected, but certainly indicative of the star being
beyond 10 \kpc.  Such stars are likely rare, and are often found
through variability studies or special spectral features \citep[see,
e.g., the carbon-star selection in][]{huxor15}.  The Besan\c{c}on
model \citep [and references therein]{robin86, robin03} indicates that
only one in fifty fields at the Galactic coordinates of {\sscma} would
have a star of comparable brightness beyond $ 10 \kpc $, and about one
in 150 beyond $ 30 \kpc $.

\section {Limits on Binarity from Radial Velocity Observations}   % Section 5

As mentioned in \S~2.2, there have been suggestions in the literature
that SS CMa might be a binary, either from the properties of a nearby
star (Star~29 in our list) or from variations of the measured RV.
In general, binarity can affect the estimated astrometric
parallax if the orbital motion of the Cepheid has a significant
component in common with the parallactic motion.  Over the short time
span of our observations ($ \sim $ 2 years, with 6-month sampling),
periods between a few months and 3 years could have a significant
impact on the measured parallax, if the orbital motion is of
sufficient amplitude and oriented appropriately.  We can rule out that
Star~29 is a binary companion on the basis of both its
spectrophotometric and astrometric parallax; even if it were
physically associated with SS CMa, its impact on the parallax would be
negligible owing to the extremely long inferred period.  On the other
hand, a spectroscopic binary companion could in principle impact the
astrometric measurement.  {\it A priori} considerations on the
likelihood of binarity as a function of period and mass ratio (see
Paper~1) suggest that the probability of a significant effect (larger
than $ 10 \muas $) is $ \sim 10\% $, but these estimates are based
primarily on binary statistics obtained for lower-mass stars
\citep[see, e.g.,][]{duquennoy91}, and thus their applicability to
massive Cepheids is uncertain.  However, in the case of Cepheids,
RV measurements can provide useful {\it direct} limits on the
possibility of binarity and its impact on the measured parallax.

\subsection {Spectroscopic Data}

We observed SS CMa between April 2013 and November 2015 using three
different echelle spectrographs: (1) the Hamilton spectrograph
\citep{vogt87} at the Shane 3~m telescope located at Lick
Observatory; (2) the Hermes spectrograph \citep{raskin11} at the
Flemish 1.2~m Mercator telescope located at the Roque de los Muchachos
Observatory on the island of La Palma, Spain; and (3) the Coralie
spectrograph \citep{queloz01} at the Swiss 1.2~m Euler telescope
located at La Silla Observatory, Chile.  Data from the Hermes and
Coralie spectrographs were reduced using dedicated pipelines.
Hamilton spectra were reduced using standard IRAF routines.  
RVs were determined by cross-correlation using a
numerical mask representative of a solar spectral type
\citep{baranne96, pepe02}. 

RVs from Hermes and Coralie were found to be compatible with each
other to within $ 10\hbox{--}20 \ms $, and no zero-point offset was applied. RVs
from the Hamilton spectrograph were brought to the Coralie/Hermes
zero-point via observations of stable standard stars (HR 4027 and one
or more of the following: HD 26161, HR 124, HR 7373, or NSV 7543)
using the velocities and zero-point offsets presented by
\citet{nidever02}.  Given the uncertainties involved with this
zero-point correction, and factoring in the intrinsic precision of the
Hamilton RVs, we estimate an uncertainty of approximately $ 200 \ms $ for
these data. The individual pipeline-estimated RV uncertainties for
Coralie and Hermes range between 20 and $ 80 \ms $.  We opt to not include
the literature RV data from \citet{joy37} and \citet{coulson85} in
this analysis, since zero-point offsets \citep[$ 1 \pm 0.5 \kms $ for
][]{coulson85} and low precision (typical uncertainties larger than 
$ 1 \kms $) dilute the precision of our new measurements, while not adding
significant information for the timescales of interest (1--2 years).

\subsection {Analysis}

To investigate a possible astrometric signal caused by binarity, we
constrain possible values of the projected semimajor axis, $ a_1 \sin
i $, for {\sscma} by modeling the RV data.  Our model consists of a
sum of a 9-harmonic Fourier series, representing the intrinsic
velocity variation during the pulsation, and a circular orbital
motion.  We adopt circular orbital motion for simplicity, and since
there is no evidence for an eccentric orbit in the available data. We
adopt a constant pulsation period of $ P = 12.3535 $~days, which
minimizes the scatter in the phased RV dataset.  We then convert the
projected semimajor axis into an astrometric term by assuming a
distance of $ 3 \kpc $.  As usual with RV information,
this represents a {\it minimum} orbital signature, corresponding to an
edge-on orbit.

\begin{figure}[ht]
\includegraphics[width=\columnwidth,bb=0 0 245 230] {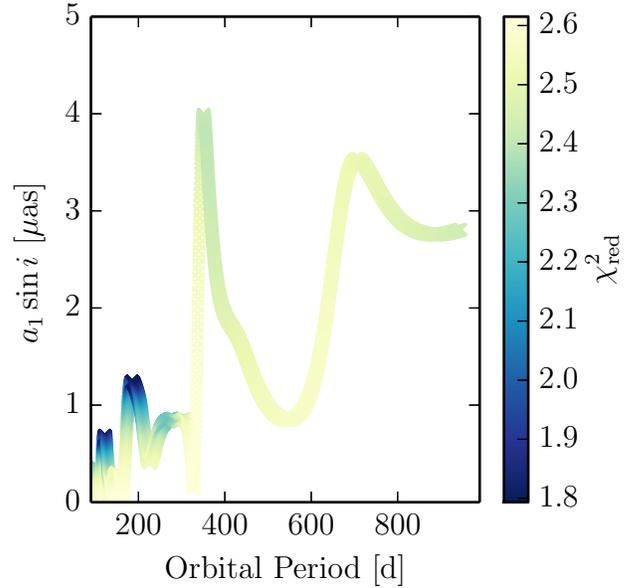}
\caption {\label{fig:rvfit} Best-fitting projected semimajor axis as a
function of period for the RV data collected for {\sscma}.
The amplitude shown here assumes that all the variation in RV
is due to orbital motion; in reality, variations in the
RV profile during pulsation are likely to contribute
substantially to the apparent RV variation, as shown by
the fact that the reduced $ \chi^2 $ remains above 1.8 throughout.
The color corresponds to the value of the reduced $ \chi^2 $, with
darker colors corresponding to better models.  The largest possible
impact on the astrometric parallax measurement is for an orbital
period $ \sim 1 $ year, and is approximately $ \pm 4 \muas $; however,
a shorter period with a much smaller astrometric impact ($ < 1.5 \muas
$) is more consistent with the RV measurements.}
\end{figure}

The time sequence of RV measurements indicates a nearly
constant systemic RV, excluding even fairly low-amplitude
deviations (above $\sim 400 \ms $), consistent with a null detection
of orbital motion to within the uncertainty of our measurements. We
estimate upper limits on astrometric signals caused by undetected
companions, noting that small variations in the pulsation RV
pattern already found in several Cepheids \citep
{anderson14b} can mimic the effect of low-mass spectroscopic
companions \citep{anderson15}.  Hence, the mere detection of
time-variable low-amplitude changes in systemic velocity is not
necessarily a clear indication of spectroscopic binarity, as the
available sampling of the RV data is insufficient to
fully separate these two possibilities.  If interpreted as orbital
motion, the variation seen in the RV data would result in
the astrometric signature shown in Figure~\ref{fig:rvfit}; darker colors
correspond to lower $ \chi^2 $.  The maximum impact is around a period
of 1 year with an orbital amplitude of $ \sim 4 \muas $, while the
lowest $ \chi^2 $ occurs for an amplitude of $ 1.5 \muas $.  In order
to account for the uncertain contribution of the possible orbital
motion, we conservatively add a term of $ 4 \muas $ in quadrature to
the nominal parallax error for SS CMa.  Monitoring of this and other
Cepheids in our program will continue, and the results will be
presented separately in greater detail (Anderson et al., in prep.).

% \clearpage

\section {Discussion}

We have presented a trigonometric parallax estimate for {\sscma}, the
first of 18 Galactic Cepheids in our program of measurements to obtain
an improved calibration of the period-luminosity relation of Cepheids
with properties comparable to those we are discovering in hosts of
Type Ia supernovae within 35 Mpc.  The parallax of {\sscma} is $ 0.348
\pm 0.038 \mas $.

Unlike the pilot case of {\syaur}, presented in Paper~1, the
availability of several bright reference stars and improvements in our
analysis procedures result in a formal uncertainty comparable with the
pre-observation expectations of 30--40 \muas.  Several of the
improvements indicated in Paper~1 have been implemented for this and
all other Cepheids in the sample.  For example, we use two shallow and
two deep scans at each epoch, we obtain serpentine scans for most
epochs, and we have developed and adopted an improved geometric
solution that removes a significant fraction of the static
(``pattern'') difference between observations obtained at different
roll angles.  We also have been able to quantify and correct for the
effect of CTE losses in the detector $ X $
direction (X-CTE), although we discovered that the second-epoch
observations were not obtained in an optimal position to minimize the
X-CTE effect.  All of these improvements will be applied to the remaining
Cepheids.

In the near term, we expect to complete the five-epoch observations
and analysis for the remaining 17 Cepheids in our sample.  If we can
achieve a comparable accuracy for the other Cepheids, we expect that
the overall characterization of the period-luminosity relation {\it
exclusively from our parallaxes} will yield an uncertainty in the
distance-scale calibration of $ \sim 2\% $, providing a powerful and
independent test of the present distance scale \citep{riess11}.  When
combined with improvements in the calibration of the distance of
NGC~4258, the contribution of the anchor distance to the calibration of
{{\hubbleconst}} is likely to drop to 1.5\%.  For a subset of 9
Cepheids, we will obtain an additional 4 epochs of observations, which
we expect to further reduce the final calibration uncertainty,
especially by improving our ability to identify and exclude outliers
and learn more about the properties of the telescope.

Additional improvements may result from a simultaneous consideration of
the data for {\it all} Cepheids.  For example, there may be
regularities in the long-term behavior of the geometric distortion of
the instrument, which we now treat as an unknown term to fit for and
subtract.  If such regularities prove amenable to a global solution,
we may be able to reduce the uncertainty in the final solution by as
much as 30\%.

In a parallel effort, we are also doubling the sample of nearby Type
Ia supernova hosts for which Cepheids are being measured and
characterized (Riess et al., in prep.).  With this two-pronged
approach, we expect to reduce the uncertainty in the {\it local}
Cepheid-based measurement of {{\hubbleconst}} by $ > 40 \% $ with
respect to \citet{riess11}.  The culmination of the Cepheid-based
efforts to refine the local measurement of {\hubbleconst} will come
with the final results from the {\it Gaia} mission (circa 2022), which
could permit, when combined with larger samples of Type Ia supernova hosts,
a measurement with accuracy of $\sim 1\% $ or better.

\acknowledgements

This project was enabled by significant assistance from a wide variety
of sources.  Merle Reinhardt, George Chapman, William Januszewski, and
Ken Sembach provided help with the {\it {\HST}} observations.  We
thank Ed Nelan, Matt Lallo, Fritz Benedict, and Barbara McArthur for
productive discussions about the behavior of the FGS.  We also thank
Leo Girardi, Alessandro Bressan, and Paola Marigo for the use of and
assistance with their Padova isochrone database, Anne Robin for
assistance with the Besan\c{c}on Galaxy Model, Eddie Schlafly and
D.~Marshall for input on the extinction along the line of sight to
{\sscma}, and Nolan Walborn for useful discussions on the
classification of hot stars.

Support for this work was provided by NASA through programs GO-12679
and GO-13101 from the Space Telescope Science Institute, which is
operated by AURA, Inc., under NASA contract NAS 5-26555.  A.V.F.'s
group at UC Berkeley is also grateful for financial assistance from
NSF grant AST-1211916, the TABASGO Foundation, and the Christopher
R. Redlich Fund.  R.I.A. acknowledges funding from the Swiss National
Science Foundation as a postdoctoral fellow.  S.C. and
A.G.R.~gratefully acknowledge support by the Munich Institute for
Astro- and Particle Physics (MIAPP) of the DFG cluster of excellence
``Origin and Structure of the Universe''.  Research at Lick
Observatory is partially supported by a generous gift from Google.

This research is based primmarily on observations with the NASA/ESA
{\it Hubble Space Telescope}, obtained at the Space Telescope Science
Institute, which is operated by AURA, Inc., under NASA contract NAS
5-26555.  Some of the data presented herein were obtained at the
W.~M.~Keck Observatory, which is operated as a scientific partnership
among the California Institute of Technology, the University of
California and the National Aeronautics and Space Administration. The
Observatory was made possible by the generous financial support of the
W.M. Keck Foundation.  Radial velocity measurements are based on
observations taken with the Coralie echelle spectrograph mounted to
the Swiss 1.2~m Euler telescope located at La Silla Observatory,
Chile, and with the Mercator Telescope, operated on the island of La
Palma by the Flemish Community, at the Spanish Observatorio del Roque
de los Muchachos of the Instituto de Astrof\'{\i}sica de Canarias.
The Euler telescope is supported by the Swiss National Science
Foundation.  Hermes is supported by the Fund for Scientific Research
of Flanders (FWO), Belgium, the Research Council of K.U.~Leuven,
Belgium, the Fonds National de la Recherche Scientifique
(F.R.S.-FNRS), Belgium, the Royal Observatory of Belgium, the
Observatoire de Gen\`eve, Switzerland, and the Th\"uringer
Landessternwarte, Tautenburg, Germany.  We thank all observers who
contributed in collecting the ground-based data used in this work, as
well as the Euler and Mercator teams for their support.

This publication makes use of data products from the {\it Wide-field
Infrared Survey Explorer (WISE)}, which is a joint project of the
University of California, Los Angeles, and the Jet Propulsion
Laboratory/California Institute of Technology, funded by NASA.  It has
also made use of the SIMBAD database, operated at CDS, Strasbourg,
France.

% \facility {Mercator1.2m}
% \facility {HST(WFC3)}

% \include {cephtable}
% \include {obstable}
% \include {startable}

\bibliographystyle{apj}
\bibliography{ss_cma_11}

\begin {table*}
\caption {Cepheids in our Sample}
\vspace {-12pt}
\begin{center}
\begin{tabular}{rccccccccc}
\hline\hline
   {Name} &    {$\log (P)$} &    {$\langle B \rangle$} &    {$\langle V \rangle$} &    {$\langle I \rangle$} & 
   {$\langle J \rangle$} &    {$\langle H \rangle$} &    {$\langle K \rangle$} & 
   {RA ($^\circ$)} &    {Dec ($^\circ$)}\\
  {} &    {(days)} &    {(mag)} &    {(mag)} &    {(mag)} &    {(mag)} &    {(mag)} &    {(mag)} &   {(J2000)} &   {(J2000)} \\
\hline
  AQ Car       &   0.990    & \phn9.785    &   8.855    &   7.870   &   7.192    &   6.743       &   6.630   & \phn$ 15.3457 $   &   $  -61.0741 $   \\
  AQ Pup       &   1.479    & \phn9.70\phn &   8.54\phn &   7.175   &   5.879    &   5.329       &   5.091   &   $  119.5920 $   &   $  -29.1301 $   \\
  CD Cyg       &   1.232    &    10.221    &   8.953    &   7.503   &   6.451    &   5.880       &   5.712   &   $  301.1107 $   &   $  +34.1123 $   \\
  DD Cas       &   0.992    &    11.111    &   9.880    &   8.580   &   7.552    &   6.952       &   6.908   &   $  359.3957 $   &   $  +62.7182 $   \\
  HW Car       &   0.964    &    10.122    &   9.125    &   8.027   &   7.258    &   6.704       &   6.596   &   $  159.8347 $   &   $  -61.1524 $   \\
  KN Cen\tablenotemark a &   1.531    &      --      &   9.86\phn &    --     &   6.27\phn &   5.92\phn    &    --     &   $  204.1537 $   &   $  -64.5583 $   \\
  RY Sco\tablenotemark a &   1.308    &      --      &   8.19\phn &    --     &    --      &   4.3\phn\phn &    --     &   $  267.7181 $   &   $  -33.7057 $   \\
   S Vul\tablenotemark a &   1.836    &      --      &   9.17\phn &    --     &   5.32\phn &   4.92\phn    &    --     &   $  297.0992 $   &   $  +27.2865 $   \\
  SS CMa       &   1.092    &    11.136    &   9.925    &   8.470   &   7.434    &   6.849       &   6.677   &   $  111.5300 $   &   $  -25.2574 $   \\
  SY Aur       &   1.006    &    10.071    &   9.066    &   7.854   &   6.899    &   6.399       &   6.391   & \phn$ 78.1634 $   &   $  +42.8318 $   \\
  SZ Cyg       &   1.179    &    10.909    &   9.430    &   7.797   &   6.573    &   5.886       &   5.746   &   $  308.2262 $   &   $  +46.6013 $   \\
  VX Per       &   1.037    &    10.459    &   9.307    &   7.995   &   7.076    &   6.517       &   6.292   & \phn$ 31.9500 $   &   $  +58.4433 $   \\
  VY Car       &   1.277    & \phn8.616    &   7.455    &   6.279   &   5.463    &   4.944       &   4.804   &   $  161.1362 $   &   $  -57.5654 $   \\
  WZ Sgr       &   1.339    & \phn9.400    &   8.017    &   6.530   &   5.402    &   4.763       &   4.565   &   $  274.2488 $   &   $  -19.0758 $   \\
   X Pup       &   1.414    & \phn9.742    &   8.515    &   7.157   &   6.180    &   5.600       &   5.430   &   $  113.2072 $   &   $  -20.9056 $   \\
  XY Car       &   1.094    &    10.510    &   9.294    &   7.950   &   6.978    &   6.405       &   6.240   &   $  165.5669 $   &   $  -64.2629 $   \\
  XZ Car       &   1.221    & \phn9.861    &   8.604    &   7.251   &   6.313    &   5.745       &   5.585   &   $  166.0561 $   &   $  -60.9799 $   \\
  YZ Car       &   1.259    & \phn9.829    &   8.709    &   7.444   &   6.492    &   5.971       &   5.808   &   $  157.0702 $   &   $  -59.3502 $   \\
   Z Sct       &   1.111    &    10.914    &   9.585    &   8.098   &   7.042    &   6.491       &   6.429   &   $  280.7386 $   & \phn$ -5.8209 $\\
\hline\hline
\end{tabular}
\end{center}
\vspace{-18pt}
\begin{center}
\begin{minipage}[t]{17cm}
\tablecomments{Cepheid data in this Table are from \citet{vanleeuwen07}, unless otherwise noted.  The photometry is used for planning purposes only;
our project will obtain {\HST} photometry for the target Cepheids.}
\par\noindent $^{\rm a}\quad${Data for this object have been obtained from the SIMBAD database \citep{wenger00}}
\end{minipage}
\end{center}
\end{table*}

\begin {table*}
\setlength{\tabcolsep}{1mm}
\renewcommand{\arraystretch}{0.88}
\tabletypesize{\scriptsize}
\tablewidth {6in}
\tablenum {2}
\caption {Properties of the Reference Stars - I.~UVIS Photometry}
\vspace{-12pt}
\begin{center}
\begin{tabular}{ccccccccc}
\hline\hline
 {Star} & {\Filter{F275W}} & {\Filter{F336W}} & {\Filter{F410M}} & {\Filter{F467M}} & {\Filter{F547M}} & {\Filter{F606W}} & {\Filter{F621M}} & {\Filter{F850LP}} \\
   {}   &   {WFC3/UVIS}    &   {WFC3/UVIS}    &   {WFC3/UVIS}    &   {WFC3/UVIS}    &   {WFC3/UVIS}    &   {WFC3/UVIS}    &   {WFC3/UVIS}    &   {WFC3/UVIS}     \\
\hline
   3 &  \nodata            &  \nodata            &  \nodata            &  15.251 $\pm$ 0.006 &  14.258 $\pm$ 0.300 &  13.850 $\pm$ 0.006 &  13.614 $\pm$ 0.003 &  \nodata            \\
   5 &  13.316 $\pm$ 0.300 &  12.768 $\pm$ 0.302 &  12.977 $\pm$ 0.301 &  \nodata            &  \nodata            &  12.345 $\pm$ 0.002 &  12.318 $\pm$ 0.001 &  12.213 $\pm$ 0.300 \\
   6 &  19.887 $\pm$ 0.090 &  18.119 $\pm$ 0.027 &  18.084 $\pm$ 0.023 &  17.498 $\pm$ 0.011 &  16.877 $\pm$ 0.003 &  16.562 $\pm$ 0.010 &  16.398 $\pm$ 0.006 &  15.513 $\pm$ 0.009 \\
   9 &  19.279 $\pm$ 0.064 &  17.520 $\pm$ 0.020 &  17.310 $\pm$ 0.015 &  16.685 $\pm$ 0.012 &     \nodata         &  15.589 $\pm$ 0.005 &  15.403 $\pm$ 0.003 &  14.251 $\pm$ 0.005 \\
  10 &  \nodata            &  \nodata            &  \nodata            &  17.336 $\pm$ 0.016 &  16.638 $\pm$ 0.006 &  16.333 $\pm$ 0.005 &  16.151 $\pm$ 0.003 &  \nodata            \\
  11 &  \nodata            &  \nodata            &  \nodata            &  17.766 $\pm$ 0.020 &  17.002 $\pm$ 0.007 &  16.662 $\pm$ 0.006 &  16.462 $\pm$ 0.004 &  \nodata            \\
  12 &  \nodata            &  \nodata            &  \nodata            &  17.853 $\pm$ 0.021 &  17.265 $\pm$ 0.008 &  16.912 $\pm$ 0.009 &  16.689 $\pm$ 0.006 &  \nodata            \\
  14 &  \nodata            &  \nodata            &  \nodata            &  12.569 $\pm$ 0.309 &  \nodata            &  12.335 $\pm$ 0.002 &  12.267 $\pm$ 0.002 &  \nodata            \\
  18 &  16.308 $\pm$ 0.014 &  15.600 $\pm$ 0.008 &  15.920 $\pm$ 0.008 &  15.630 $\pm$ 0.007 &  15.078 $\pm$ 0.004 &  14.857 $\pm$ 0.005 &  14.724 $\pm$ 0.002 &  13.872 $\pm$ 0.004 \\
  21 &  19.246 $\pm$ 0.066 &  17.989 $\pm$ 0.025 &  17.361 $\pm$ 0.016 &  16.875 $\pm$ 0.011 &  16.328 $\pm$ 0.005 &  16.061 $\pm$ 0.007 &  15.905 $\pm$ 0.008 &  14.954 $\pm$ 0.007 \\
  23 &  20.202 $\pm$ 0.133 &  18.824 $\pm$ 0.039 &  18.762 $\pm$ 0.032 &  18.149 $\pm$ 0.021 &  17.456 $\pm$ 0.009 &  17.081 $\pm$ 0.010 &  16.902 $\pm$ 0.009 &  15.887 $\pm$ 0.011 \\
  25 &  18.282 $\pm$ 0.038 &  16.498 $\pm$ 0.012 &  16.250 $\pm$ 0.009 &  \nodata            &  \nodata            &  14.725 $\pm$ 0.004 &  14.539 $\pm$ 0.007 &  13.532 $\pm$ 0.300 \\
  26 &  16.557 $\pm$ 0.016 &  15.474 $\pm$ 0.007 &  14.790 $\pm$ 0.004 &  14.689 $\pm$ 0.300 &  14.000 $\pm$ 0.301 &  13.716 $\pm$ 0.003 &  13.565 $\pm$ 0.005 &  12.544 $\pm$ 0.300 \\
  29 &  18.110 $\pm$ 0.036 &  16.429 $\pm$ 0.012 &  16.524 $\pm$ 0.011 &  16.003 $\pm$ 0.005 &  15.571 $\pm$ 0.003 &  15.333 $\pm$ 0.005 &  15.210 $\pm$ 0.004 &  14.658 $\pm$ 0.006 \\
  31 &  18.611 $\pm$ 0.044 &  17.005 $\pm$ 0.015 &  16.929 $\pm$ 0.013 &  16.375 $\pm$ 0.009 &  15.764 $\pm$ 0.004 &  15.469 $\pm$ 0.005 &  15.292 $\pm$ 0.008 &  14.369 $\pm$ 0.005 \\
  37 &  17.374 $\pm$ 0.024 &  15.634 $\pm$ 0.008 &  15.071 $\pm$ 0.300 &  13.649 $\pm$ 0.301 &  \nodata            &  12.237 $\pm$ 0.003 &  11.988 $\pm$ 0.005 &  \nodata            \\
  38 &  19.433 $\pm$ 0.074 &  17.588 $\pm$ 0.020 &  17.368 $\pm$ 0.016 &  16.674 $\pm$ 0.010 &  15.692 $\pm$ 0.004 &  15.552 $\pm$ 0.007 &  15.355 $\pm$ 0.009 &  14.198 $\pm$ 0.005 \\
  39 &  \nodata            &  \nodata            &  \nodata            &  17.646 $\pm$ 0.017 &  16.980 $\pm$ 0.007 &  16.691 $\pm$ 0.012 &  16.481 $\pm$ 0.009 &  \nodata            \\
  43 &  \nodata            &  \nodata            &  \nodata            &  17.291 $\pm$ 0.014 &  15.921 $\pm$ 0.004 &  15.260 $\pm$ 0.004 &  14.972 $\pm$ 0.006 &  \nodata            \\
  44 &  \nodata            &  \nodata            &  \nodata            &  15.974 $\pm$ 0.007 &  15.551 $\pm$ 0.003 &  15.339 $\pm$ 0.005 &  15.202 $\pm$ 0.009 &  \nodata            \\
  45 &  \nodata            &  \nodata            &  \nodata            &  17.246 $\pm$ 0.014 &  16.137 $\pm$ 0.005 &  15.588 $\pm$ 0.004 &  15.314 $\pm$ 0.008 &  \nodata            \\
  66 &  20.564 $\pm$ 0.160 &  19.048 $\pm$ 0.042 &  18.896 $\pm$ 0.034 &  18.180 $\pm$ 0.022 &  17.464 $\pm$ 0.009 &  17.085 $\pm$ 0.008 &  16.868 $\pm$ 0.009 &  15.683 $\pm$ 0.010 \\
  n1 &  \nodata            &  \nodata            &  \nodata            &  18.424 $\pm$ 0.028 &  17.785 $\pm$ 0.011 &  \nodata            &  \nodata            &  \nodata            \\
  n2 &  20.976 $\pm$ 0.419 &  19.646 $\pm$ 0.061 &  19.513 $\pm$ 0.049 &  18.884 $\pm$ 0.021 &  18.073 $\pm$ 0.009 &  17.673 $\pm$ 0.010 &  17.469 $\pm$ 0.006 &  16.315 $\pm$ 0.013 \\
  n4 &  23.488 $\pm$ 1.397 &  20.552 $\pm$ 0.108 &  20.027 $\pm$ 0.064 &  18.979 $\pm$ 0.033 &  17.938 $\pm$ 0.011 &  17.391 $\pm$ 0.010 &  17.148 $\pm$ 0.009 &  15.518 $\pm$ 0.009 \\
  n7 &  20.451 $\pm$ 0.166 &  19.202 $\pm$ 0.047 &  \nodata            &  18.366 $\pm$ 0.024 &  17.535 $\pm$ 0.301 &  17.264 $\pm$ 0.010 &  17.053 $\pm$ 0.014 &  15.972 $\pm$ 0.011 \\
 n10 &  23.207 $\pm$ 1.152 &  20.534 $\pm$ 0.104 &  19.766 $\pm$ 0.056 &  18.645 $\pm$ 0.027 &  17.626 $\pm$ 0.010 &  17.109 $\pm$ 0.008 &  16.847 $\pm$ 0.006 &  15.174 $\pm$ 0.008 \\
 n11 &  21.637 $\pm$ 0.790 &  20.674 $\pm$ 0.118 &  20.065 $\pm$ 0.065 &  19.026 $\pm$ 0.038 &  18.012 $\pm$ 0.017 &  17.484 $\pm$ 0.010 &  17.224 $\pm$ 0.004 &  15.629 $\pm$ 0.009 \\
\hline
\end{tabular}
\end{center}
\vspace{-18pt}
\begin{center}
\tablecomments{All {\HST} photometric measurements are in the Vega system and have been obtained from data for this Project.  
Magnitudes in {\Filter{F606W}} and {\Filter{F621M}} are from observations obtained with spatial scanning data; 
other measurements are from $ 2 \times 2 $ binned data.}
\end{center}
\end{table*}    

\begin {table*}
\setlength{\tabcolsep}{1mm}
\renewcommand{\arraystretch}{0.88}
\tabletypesize{\scriptsize}
\tablewidth {6in}
\tablenum {2}
\caption {Properties of the Reference Stars - II.~IR Photometry}
\vspace{-12pt}
\begin{center}
\begin{tabular}{ccccccc}
\hline\hline
    {Star} & {\Filter{J}} & {\Filter{F160W}} & {\Filter{H}}  & {\Filter{K}}    &  {Channel 1} &  {Channel 2}  \\
    {}     &   {2MASS}    &    {WFC3/IR}     &     {2MASS}   &     {2MASS}     &     {WISE}   &     {WISE}    \\
\hline
   3 &       11.261 $\pm$ 0.022 &   \nodata           &       10.642 $\pm$ 0.021 &       10.439 $\pm$ 0.023 &  10.282  $\pm$ 0.026 &  10.322  $\pm$ 0.026 \\
   5 &       11.902 $\pm$ 0.022 &   \nodata           &       11.813 $\pm$ 0.022 &       11.802 $\pm$ 0.024 &   \nodata            &   \nodata            \\
   6 &       14.882 $\pm$ 0.042 &   \nodata           &       14.375 $\pm$ 0.064 &       14.478 $\pm$ 0.094 &   \nodata            &   \nodata            \\
   9 &       13.575 $\pm$ 0.026 &   \nodata           &       13.099 $\pm$ 0.021 &       12.996 $\pm$ 0.036 &   \nodata            &   \nodata            \\
  10 &       14.435 $\pm$ 0.027 &   \nodata           &       13.960 $\pm$ 0.043 &       13.976 $\pm$ 0.058 &  13.880  $\pm$ 0.030 &  13.957  $\pm$ 0.047 \\
  11 &       14.573 $\pm$ 0.026 &   \nodata           &       14.173 $\pm$ 0.035 &       14.014 $\pm$ 0.063 &  13.829  $\pm$ 0.031 &  14.040  $\pm$ 0.053 \\
  12 &       14.959 $\pm$ 0.059 &   \nodata           &       14.374 $\pm$ 0.065 &   \nodata                &  13.960  $\pm$ 0.032 &  14.049  $\pm$ 0.051 \\
  14 &       11.646 $\pm$ 0.024 &   \nodata           &       11.475 $\pm$ 0.021 &       11.455 $\pm$ 0.024 &  11.358  $\pm$ 0.023 &   \nodata            \\
  18 &       13.365 $\pm$ 0.022 &   \nodata           &       13.162 $\pm$ 0.034 &       13.093 $\pm$ 0.034 &   \nodata            &   \nodata            \\
  21 &       14.342 $\pm$ 0.024 &  14.125 $\pm$ 0.017 &       14.056 $\pm$ 0.047 &       13.877 $\pm$ 0.059 &   \nodata            &   \nodata            \\
  23 &   \nodata                &  14.840 $\pm$ 0.023 &   \nodata                &   \nodata                &   \nodata            &   \nodata            \\
  25 &       12.914 $\pm$ 0.024 &  12.581 $\pm$ 0.008 &       12.514 $\pm$ 0.024 &       12.417 $\pm$ 0.029 &   \nodata            &   \nodata            \\
  26 &       12.110 $\pm$ 0.021 &  11.935 $\pm$ 0.006 &       11.922 $\pm$ 0.024 &       11.761 $\pm$ 0.023 &   \nodata            &   \nodata            \\
  29 &   \nodata                &   \nodata           &   \nodata                &   \nodata                &   \nodata            &   \nodata            \\
  31 &       13.759 $\pm$ 0.021 &  13.417 $\pm$ 0.012 &       13.326 $\pm$ 0.022 &       13.282 $\pm$ 0.036 &   \nodata            &   \nodata            \\
  37 &  {\phn9.453} $\pm$ 0.022 &   \nodata           &  {\phn8.703} $\pm$ 0.061 &  {\phn8.481} $\pm$ 0.019 &   \nodata            &   \nodata            \\
  38 &       13.402 $\pm$ 0.024 &  12.993 $\pm$ 0.010 &       12.876 $\pm$ 0.022 &       12.782 $\pm$ 0.027 &   \nodata            &   \nodata            \\
  39 &       14.910 $\pm$ 0.047 &   \nodata           &       14.395 $\pm$ 0.044 &       14.219 $\pm$ 0.073 &   \nodata            &   \nodata            \\
  43 &       11.741 $\pm$ 0.030 &   \nodata           &       10.762 $\pm$ 0.028 &       10.492 $\pm$ 0.024 &   \nodata            &  10.336  $\pm$ 0.022 \\
  44 &   \nodata                &  14.449 $\pm$ 0.015 &   \nodata                &   \nodata                &   \nodata            &   \nodata            \\
  45 &       12.555 $\pm$ 0.022 &   \nodata           &       11.749 $\pm$ 0.026 &       11.486 $\pm$ 0.021 &   \nodata            &  11.406  $\pm$ 0.023 \\
  66 &   \nodata                &  14.484 $\pm$ 0.020 &   \nodata                &   \nodata                &   \nodata            &   \nodata            \\
  n1 &       15.631 $\pm$ 0.062 &   \nodata           &       15.543 $\pm$ 0.118 &       15.515 $\pm$ 0.216 &  15.014  $\pm$ 0.042 &   \nodata            \\
  n2 &       15.663 $\pm$ 0.055 &   \nodata           &       15.020 $\pm$ 0.059 &   \nodata                &   \nodata            &   \nodata            \\
  n4 &       14.480 $\pm$ 0.036 &  13.879 $\pm$ 0.015 &       13.704 $\pm$ 0.026 &       13.475 $\pm$ 0.035 &   \nodata            &   \nodata            \\
  n7 &       15.319 $\pm$ 0.037 &  14.901 $\pm$ 0.024 &       14.802 $\pm$ 0.047 &   \nodata                &   \nodata            &   \nodata            \\
 n10 &       14.145 $\pm$ 0.035 &   \nodata           &       13.400 $\pm$ 0.029 &       13.119 $\pm$ 0.033 &   \nodata            &   \nodata            \\
 n11 &       14.671 $\pm$ 0.027 &   \nodata           &       13.888 $\pm$ 0.038 &       13.780 $\pm$ 0.046 &   \nodata            &   \nodata            \\
\hline\hline
\end{tabular}
\end{center}
\vspace{-18pt}
\begin{center}
\tablecomments{All {\HST} photometric measurements are in the Vega system and have been obtained from data for this Project. 
Photometry obtained from the 2MASS Survey \citep{skrutskie06} and from {\it WISE} \citep{wright10} are in their respective systems.}
\end{center}
\end{table*}    

\begin {table*}
\setlength{\tabcolsep}{1mm}
\renewcommand{\arraystretch}{0.88}
\tabletypesize{\scriptsize}
\tablewidth {6in}
\tablenum {2}
\tabletypesize{\scriptsize}
\caption {Properties of the Reference Stars - III.  Spectra, Classification, and Astrometry}
\vspace{-12pt}
\begin{center}
\begin{tabular}{ccccccccc}
\hline\hline
  {Star} & {RA ($^\circ$} & {Dec ($^\circ$)} & {Spectrum} & {Class\tablenotemark a} & {Quality\tablenotemark b} & {$T_{\rm eff}$\tablenotemark c} & {$ \pi_{\rm full} $\tablenotemark d} & {$ \pi_{\rm astro}$\tablenotemark e}\\
 {}  & {(J2000)}    & {(J2000)}  & {Source}   & {}   & {}     & {(K)~~}      &   {(mas)}     &  {(mas)}\\
\hline
    3   &  111.49716   & -25.23251    &    Gemini   &  K0~IV       &   F    & {\phn 5340\phn}  $\pm$ {\phn160} &   0.414 $\pm$ 0.043  &  0.166 $\pm$ 0.110 \\
    5   &  111.51980   & -25.24066    &    Gemini   &  A7~V        &   G    & {\phn 7920\phn}  $\pm$ {\phn250} &   1.185 $\pm$ 0.049  &  1.197 $\pm$ 0.053 \\
    6   &  111.51377   & -25.26266    &    Gemini   &  G5~IV       &   VG   & {\phn 5730\phn}  $\pm$ {\phn110} &   0.746 $\pm$ 0.045  &  0.758 $\pm$ 0.051 \\
    9   &  111.53022   & -25.23712    &    Lick     &  G0~IV--V    &   VG   & {\phn 6180\phn}  $\pm$ {\phn180} & \nodata              & \nodata            \\
   10   &  111.54177   & -25.23063    &    Gemini   &  F7~IV       &   VG   & {\phn 6550\phn}  $\pm$ {\phn200} &   0.448 $\pm$ 0.096  &  0.060 $\pm$ 0.260 \\
   11   &  111.52409   & -25.22336    &    Lick     &  F8~III      &   F    & {\phn 6420\phn}  $\pm$ {\phn160} & \nodata              & \nodata            \\
   12   &  111.51990   & -25.22623    &    Gemini   &  K5~IV       &   VG   & {\phn 4600\phn}  $\pm$ {\phn260} & \nodata              & \nodata            \\
   14   &  111.51993   & -25.23608    &    Lick     &  F5~IV--V    &   VG   & {\phn 6700\phn}  $\pm$ {\phn130} &   1.747 $\pm$ 0.061  &  1.766 $\pm$ 0.063 \\
   18   &  111.51415   & -25.25188    &    Lick     &  B2~V        &   F    &     {21000\phn}  $\pm$    {1000} &   0.281 $\pm$ 0.023  &  0.339 $\pm$ 0.041 \\
   21   &  111.53000   & -25.26841    &    Gemini   &  A4~III--IV  &   VG   & {\phn 8400\phn}  $\pm$ {\phn340} &   0.333 $\pm$ 0.027  &  0.297 $\pm$ 0.036 \\
   23   &  111.53011   & -25.26374    &    Gemini   &  F8~V        &   VG   & {\phn 6420\phn}  $\pm$ {\phn160} &   0.398 $\pm$ 0.036  &  0.386 $\pm$ 0.040 \\
   25   &  111.54306   & -25.27473    &    Lick     &  G5~IV--V    &   G    & {\phn 5730\phn}  $\pm$ {\phn110} &   0.688 $\pm$ 0.037  &  0.683 $\pm$ 0.040 \\
   26   &  111.52272   & -25.28065    &    Lick     &  A2~III--IV  &   VG   & {\phn 8680\phn}  $\pm$ {\phn360} & \nodata              & \nodata            \\
   29   &  111.53344   & -25.25502    &    Gemini   &  G6~V        &   VG   & {\phn 5690\phn}  $\pm$ {\phn130} &   1.012 $\pm$ 0.038  &  1.023 $\pm$ 0.041 \\
   31   &  111.54654   & -25.27182    &    Gemini   &  F8~IV       &   VG   & {\phn 6420\phn}  $\pm$ {\phn160} &   0.636 $\pm$ 0.038  &  0.644 $\pm$ 0.042 \\
   37   &  111.54825   & -25.26604    &    Gemini   &  G2~III      &   VG   & {\phn 5620\phn}  $\pm$ {\phn180} & \nodata              & \nodata            \\
   38   &  111.53878   & -25.26636    &    Lick     &  G0~V        &   VG   & {\phn 6180\phn}  $\pm$ {\phn180} &   0.661 $\pm$ 0.031  &  0.699 $\pm$ 0.035 \\
   39   &  111.56396   & -25.28153    &    Lick     &  G0~V        &   F    & {\phn 5920\phn}  $\pm$ {\phn180} &   0.730 $\pm$ 0.092  &  0.789 $\pm$ 0.117 \\
   43   &  111.52175   & -25.28641    &    Keck     &  K6~Iab      &   G    & {\phn 5500\phn}  $\pm$ {\phn260} &   0.032 $\pm$ 0.008  &  0.018 $\pm$ 0.064 \\
   44   &  111.52294   & -25.28780    &    Lick     &  G2~IV--V    &   G    & {\phn 5800\phn}  $\pm$ {\phn180} &   0.797 $\pm$ 0.037  &  0.799 $\pm$ 0.039 \\
   45   &  111.53057   & -25.28724    &    Lick     &  F8~II--III  &   F    & {\phn 6300\phn}  $\pm$ {\phn130} &   0.257 $\pm$ 0.022  &  0.336 $\pm$ 0.053 \\
   66   &  111.55038   & -25.26568    &    Lick     &  G2~III      &   G    & {\phn 5620\phn}  $\pm$ {\phn180} &   0.366 $\pm$ 0.040  &  0.337 $\pm$ 0.071 \\
   n1   &  111.50797   & -25.23456    &    Lick     &  B9\tablenotemark{f} & P &             \nodata           &   0.548 $\pm$ 0.063  &  0.548 $\pm$ 0.074 \\
   n2   &  111.51711   & -25.26131    &    Keck     &  F7~III--IV  &   G    & {\phn 6480\phn}  $\pm$ {\phn200} & \nodata              & \nodata            \\
   n4   &  111.52766   & -25.26668    &    Keck     &  G5\tablenotemark{f} & F & {\phn 5660\phn} $\pm$ {\phn110} & 0.240 $\pm$ 0.024  &  0.124 $\pm$ 0.043 \\
   n7   &  111.54059   & -25.26129    &    Keck     &  F7~V        &   G    & {\phn 6480\phn}  $\pm$ {\phn200} & \nodata              & \nodata            \\
  n10   &  111.55420   & -25.25736    &    Keck     &  G5\tablenotemark{f} & G & {\phn 5660\phn} $\pm$ {\phn110} & \nodata            & \nodata            \\
  n11   &  111.50804   & -25.25449    &    Keck     &  G9~IV--V    &   G    & {\phn 5390\phn}  $\pm$ {\phn160} &  0.331  $\pm$ 0.036  &  0.376 $\pm$ 0.060 \\
\hline\hline
\end{tabular}
\end{center}
\vspace{-18pt}
\begin{center}
\begin{minipage}[t]{17cm}
\par\noindent $^{\rm a}\quad${Spectral classification based on MKCLASS Version 1.7 \citep{gray14} (see text for details) unless otherwise noted}
\par\noindent $^{\rm b}\quad${Classification quality from MKCLASS; VG=Very good, G=Good, F=Fair, P=Poor}
\par\noindent $^{\rm c}\quad${Temperature and uncertainty from the spectroscopic analysis}
\par\noindent $^{\rm d}\quad${Parallax estimate based on combined spectrophotometric and astrometric data}
\par\noindent $^{\rm e}\quad${Parallax estimate based purely on astrometric data for each star; spectrophotometric information is retained for all other stars.  See Section~4.2 for details.}
\par\noindent $^{\rm f}\quad${Classification and quality from match to model spectra; luminosity class not available}
\end{minipage}
\end{center}
\end{table*}

\begin{table*}
\tabletypesize{\scriptsize}
\renewcommand{\arraystretch}{0.80}
\tablenum{3}
\caption {Spatial Scanning Observations Used in this Paper}
\begin{tabular}{cccccccccr}
\hline
\hline
    {Date} &     {Rootname} &     {Program ID}  &     {EXPSTART} &
    { Filter} &     {Exp.~time} &     {Scan rate} &     {Scan length} &     {Number} &    {$ X$ pos targ}\\
   {} &     {} &     {} &     {(MJD)} &     {} &     { (seconds)} & 
      {($\arcsec \, \rm s^{-1} $)} &     {($\arcsec$)} &     {of legs} &     {($\arcsec$)}\\
\hline
\multicolumn{5}{l}{\bf Observations for SS CMa}\\
\hline
2012-10-23 & ibzc04kjq & 12879 & 56223.905323 & \Filter{F606W} & 350.0 &  0.330 & 115.5 & 1 & $  -3.00  $ \\ 
2012-10-23 & ibzc04klq & 12879 & 56223.911133 & \Filter{F621M} & 350.0 &  0.330 & 115.5 & 1 & $  -3.00  $ \\ 
2012-10-23 & ibzc04knq & 12879 & 56223.916712 & \Filter{F621M} & 350.0 &  0.330 & 115.5 & 1 & $  -3.00  $ \\ 
2012-10-23 & ibzc04kpq & 12879 & 56223.922592 & \Filter{F606W} & 350.0 &  0.330 & 115.5 & 1 & $  -3.00  $ \\ 
2012-10-23 & ibzc04krq & 12879 & 56223.928899 & \Filter{F606W} & 350.0 &  1.505 & 526.7 & 3 & $ -10.00  $ \\ 
\hline
2013-04-18 & ibzc15ntq & 12879 & 56400.332005 & \Filter{F606W} & 348.0 &  0.330 & 114.8 & 1 & $  -3.00  $ \\ 
2013-04-18 & ibzc15nvq & 12879 & 56400.337792 & \Filter{F621M} & 348.0 &  0.330 & 114.8 & 1 & $  -3.00  $ \\ 
2013-04-18 & ibzc15nxq & 12879 & 56400.343347 & \Filter{F621M} & 348.0 &  0.330 & 114.8 & 1 & $  -3.00  $ \\ 
2013-04-18 & ibzc15o0q & 12879 & 56400.349204 & \Filter{F606W} & 348.0 &  0.330 & 114.8 & 1 & $  -3.00  $ \\ 
2013-04-18 & ibzc15o2q & 12879 & 56400.355489 & \Filter{F606W} & 348.0 &  1.505 & 523.7 & 3 & $ -10.00  $ \\ 
\hline
2013-10-22 & ic8z04haq & 13344 & 56587.580458 & \Filter{F606W} & 348.0 &  0.330 & 114.8 & 1 & $  -3.00  $ \\ 
2013-10-22 & ic8z04hcq & 13344 & 56587.586246 & \Filter{F621M} & 348.0 &  0.330 & 114.8 & 1 & $  -3.00  $ \\ 
2013-10-22 & ic8z04heq & 13344 & 56587.591801 & \Filter{F621M} & 348.0 &  0.330 & 114.8 & 1 & $  -3.00  $ \\ 
2013-10-22 & ic8z04hgq & 13344 & 56587.597657 & \Filter{F606W} & 348.0 &  0.330 & 114.8 & 1 & $  -3.00  $ \\ 
2013-10-22 & ic8z04hiq & 13344 & 56587.603942 & \Filter{F606W} & 348.0 &  1.505 & 523.7 & 3 & $ -10.00  $ \\ 
\hline
2014-04-16 & ic8z15o4q & 13344 & 56763.310662 & \Filter{F606W} & 348.0 &  0.330 & 114.8 & 1 & $  +3.00  $ \\ 
2014-04-16 & ic8z15o6q & 13344 & 56763.316449 & \Filter{F621M} & 348.0 &  0.330 & 114.8 & 1 & $  +3.00  $ \\ 
2014-04-16 & ic8z15o8q & 13344 & 56763.322005 & \Filter{F621M} & 348.0 &  0.330 & 114.8 & 1 & $  +3.00  $ \\ 
2014-04-16 & ic8z15oaq & 13344 & 56763.327862 & \Filter{F606W} & 348.0 &  0.330 & 114.8 & 1 & $  +3.00  $ \\ 
2014-04-16 & ic8z15ocq & 13344 & 56763.334146 & \Filter{F606W} & 348.0 &  1.505 & 523.7 & 3 & $  +2.00  $ \\ 
\hline
2014-10-23 & icir03ixq & 13678 & 56953.729151 & \Filter{F606W} & 348.0 &  0.330 & 114.8 & 1 & $  -3.00  $ \\ 
2014-10-23 & icir03j3q & 13678 & 56953.746350 & \Filter{F606W} & 348.0 &  0.330 & 114.8 & 1 & $  -3.00  $ \\ 
2014-10-23 & icir03izq & 13678 & 56953.734938 & \Filter{F621M} & 348.0 &  0.330 & 114.8 & 1 & $  -3.00  $ \\ 
2014-10-23 & icir03j1q & 13678 & 56953.740494 & \Filter{F621M} & 348.0 &  0.330 & 114.8 & 1 & $  -3.00  $ \\ 
\hline
\multicolumn{5}{l}{\bf Observations for M48}\\
\hline
2014-09-25 & icmp04ptq & 13929 & 56925.069671 & \Filter{F673N} & 348.0 & 0.400 & 139.2 & 1 & $   -6.90 $ \\
2014-09-25 & icmp04pvq & 13929 & 56925.075678 & \Filter{F673N} & 348.0 & 0.400 & 139.2 & 1 & $  -33.10 $ \\
2014-09-25 & icmp04pyq & 13929 & 56925.081812 & \Filter{F673N} & 348.0 & 0.400 & 139.2 & 1 & $  +26.20 $ \\
2014-09-25 & icmp04q0q & 13929 & 56925.087738 & \Filter{F673N} & 348.0 & 0.400 & 139.2 & 1 & $  +14.40 $ \\
2014-09-25 & icmp04q2q & 13929 & 56925.131662 & \Filter{F673N} & 348.0 & 0.400 & 139.2 & 1 & $  +43.50 $ \\
2014-09-25 & icmp06qnq & 13929 & 56925.274092 & \Filter{F673N} & 348.0 & 0.400 & 139.2 & 1 & $   +8.50 $ \\
2014-09-25 & icmp06qpq & 13929 & 56925.280227 & \Filter{F673N} & 348.0 & 0.400 & 139.2 & 1 & $  +57.50 $ \\
2014-09-25 & icmp06qrq & 13929 & 56925.286199 & \Filter{F673N} & 348.0 & 0.400 & 139.2 & 1 & $  -12.00 $ \\
2014-09-25 & icmp06qtq & 13929 & 56925.337796 & \Filter{F673N} & 348.0 & 0.400 & 139.2 & 1 & $  +17.40 $ \\
2014-09-25 & icmp06qvq & 13929 & 56925.343907 & \Filter{F673N} & 348.0 & 0.400 & 139.2 & 1 & $  -24.00 $ \\
2014-09-28 & icmp05xlq & 13929 & 56928.863641 & \Filter{F673N} & 348.0 & 0.400 & 139.2 & 1 & $  -21.20 $ \\
2014-09-28 & icmp05xnq & 13929 & 56928.869741 & \Filter{F673N} & 348.0 & 0.400 & 139.2 & 1 & $  +36.30 $ \\
2014-09-28 & icmp05xpq & 13929 & 56928.907391 & \Filter{F673N} & 348.0 & 0.400 & 139.2 & 1 & $  +16.30 $ \\
2014-09-28 & icmp05xrq & 13929 & 56928.913410 & \Filter{F673N} & 348.0 & 0.400 & 139.2 & 1 & $  -47.20 $ \\
2014-09-28 & icmp05xtq & 13929 & 56928.919486 & \Filter{F673N} & 348.0 & 0.400 & 139.2 & 1 & $   +3.30 $ \\
2014-09-29 & icmp01gmq & 13929 & 56929.974255 & \Filter{F621M} & 348.0 & 0.400 & 139.2 & 1 & $   -6.90 $ \\
2014-09-29 & icmp01goq & 13929 & 56929.980261 & \Filter{F621M} & 348.0 & 0.400 & 139.2 & 1 & $  -33.10 $ \\
2014-09-29 & icmp01gqq & 13929 & 56929.986396 & \Filter{F621M} & 348.0 & 0.400 & 139.2 & 1 & $  +26.20 $ \\
2014-09-29 & icmp01gsq & 13929 & 56929.992322 & \Filter{F621M} & 348.0 & 0.400 & 139.2 & 1 & $  +14.40 $ \\
2014-09-29 & icmp01guq & 13929 & 56929.998340 & \Filter{F621M} & 348.0 & 0.400 & 139.2 & 1 & $  +43.50 $ \\
2014-09-30 & icmp03h5q & 13929 & 56930.039567 & \Filter{F621M} & 348.0 & 0.400 & 139.2 & 1 & $   +8.50 $ \\
2014-09-30 & icmp03haq & 13929 & 56930.045701 & \Filter{F621M} & 348.0 & 0.400 & 139.2 & 1 & $  +57.50 $ \\
2014-09-30 & icmp03hcq & 13929 & 56930.051674 & \Filter{F621M} & 348.0 & 0.400 & 139.2 & 1 & $  -12.00 $ \\
2014-09-30 & icmp03hfq & 13929 & 56930.057819 & \Filter{F621M} & 348.0 & 0.400 & 139.2 & 1 & $  +17.40 $ \\
2014-09-30 & icmp03hhq & 13929 & 56930.063931 & \Filter{F621M} & 348.0 & 0.400 & 139.2 & 1 & $  -24.00 $ \\
2014-10-01 & icmp02odq & 13929 & 56931.174394 & \Filter{F621M} & 348.0 & 0.400 & 139.2 & 1 & $  -21.20 $ \\
2014-10-01 & icmp02ofq & 13929 & 56931.180493 & \Filter{F621M} & 348.0 & 0.400 & 139.2 & 1 & $  +36.30 $ \\
2014-10-01 & icmp02ohq & 13929 & 56931.186662 & \Filter{F621M} & 348.0 & 0.400 & 139.2 & 1 & $  +16.30 $ \\
2014-10-01 & icmp02ojq & 13929 & 56931.192681 & \Filter{F621M} & 348.0 & 0.400 & 139.2 & 1 & $  -47.20 $ \\
2014-10-01 & icmp02opq & 13929 & 56931.236743 & \Filter{F621M} & 348.0 & 0.400 & 139.2 & 1 & $   +3.30 $ \\
\hline\hline
\end{tabular}
\end{table*}

\begin{table*}
\tabletypesize{\scriptsize}
\renewcommand{\arraystretch}{0.80}
\tablenum{3}
\caption {Spatial Scanning Observations Used in this Paper (continued)}
\begin{tabular}{cccccccccr}
\hline
\hline
    {Date} &     {Rootname} &     {Program ID}  &     {EXPSTART} &
    { Filter} &     {Exp.~time} &     {Scan rate} &     {Scan length} &     {Number} &    {$ X$ pos targ}\\
   {} &     {} &     {} &     {(MJD)} &     {} &     { (seconds)} & 
      {($\arcsec \, \rm s^{-1} $)} &     {($\arcsec$)} &     {of legs} &     {($\arcsec$)}\\
\hline
\multicolumn{5}{l}{\bf Observations for M67}\\
\hline
2014-11-08 & icmp07m8q & 13929 & 56969.071721 & \Filter{F606W} & 348.0 & 0.400 & 139.2 & 1 & $ -30.34 $ \\
2014-11-08 & icmp07maq & 13929 & 56969.077670 & \Filter{F606W} & 348.0 & 0.400 & 139.2 & 1 & $ -14.35 $ \\
2014-11-08 & icmp07mdq & 13929 & 56969.083758 & \Filter{F606W} & 348.0 & 0.400 & 139.2 & 1 & $ +30.88 $ \\
2014-11-08 & icmp07mfq & 13929 & 56969.132601 & \Filter{F606W} & 348.0 & 0.400 & 139.2 & 1 & $ +36.42 $ \\
2014-11-08 & icmp07mhq & 13929 & 56969.138955 & \Filter{F606W} & 348.0 & 0.400 & 139.2 & 1 & $ +39.60 $ \\
2014-11-08 & icmp08nyq & 13929 & 56969.524129 & \Filter{F606W} & 350.0 & 0.400 & 140.0 & 1 & $  +7.05 $ \\
2014-11-08 & icmp08o0q & 13929 & 56969.530055 & \Filter{F606W} & 350.0 & 0.400 & 140.0 & 1 & $  -2.88 $ \\
2014-11-08 & icmp08o2q & 13929 & 56969.536003 & \Filter{F606W} & 350.0 & 0.400 & 140.0 & 1 & $  +8.96 $ \\
2014-11-08 & icmp08o4s & 13929 & 56969.541988 & \Filter{F606W} & 350.0 & 0.400 & 140.0 & 1 & $ +26.71 $ \\
2014-11-08 & icmp08o6s & 13929 & 56969.547902 & \Filter{F606W} & 350.0 & 0.400 & 140.0 & 1 & $ +18.54 $ \\
2014-11-08 & icmp10q6q & 13929 & 56969.922312 & \Filter{F606W} & 350.0 & 0.400 & 140.0 & 1 & $  +3.90 $ \\
2014-11-08 & icmp10q8q & 13929 & 56969.928295 & \Filter{F606W} & 350.0 & 0.400 & 140.0 & 1 & $  +0.28 $ \\
2014-11-08 & icmp10qaq & 13929 & 56969.934302 & \Filter{F606W} & 350.0 & 0.400 & 140.0 & 1 & $  +3.89 $ \\
2014-11-08 & icmp10qcq & 13929 & 56969.940321 & \Filter{F606W} & 350.0 & 0.400 & 140.0 & 1 & $ -16.75 $ \\
2014-11-08 & icmp10qeq & 13929 & 56969.946281 & \Filter{F606W} & 350.0 & 0.400 & 140.0 & 1 & $ -22.13 $ \\
2014-11-09 & icmp09vkq & 13929 & 56970.851154 & \Filter{F606W} & 350.0 & 0.400 & 140.0 & 1 & $  -5.14 $ \\
2014-11-09 & icmp09vmq & 13929 & 56970.857022 & \Filter{F606W} & 350.0 & 0.400 & 140.0 & 1 & $ +13.52 $ \\
2014-11-09 & icmp09voq & 13929 & 56970.862890 & \Filter{F606W} & 350.0 & 0.400 & 140.0 & 1 & $  -8.37 $ \\
2014-11-09 & icmp09vqq & 13929 & 56970.868886 & \Filter{F606W} & 350.0 & 0.400 & 140.0 & 1 & $ +15.69 $ \\
2014-11-09 & icmp09vsq & 13929 & 56970.874777 & \Filter{F606W} & 350.0 & 0.400 & 140.0 & 1 & $  +2.15 $ \\
\hline
\hline
\end{tabular}
\end{table*}

\end{document}